\documentclass[lettersize,journal, onecolumn]{IEEEtran}  
\usepackage{amsmath,amsfonts}
\usepackage{algorithm}
\usepackage{algpseudocode}
\usepackage{array}
\usepackage[caption=true,font=small,labelfont=sf,textfont=sf]{subfig}
\usepackage{textcomp}
\usepackage{stfloats}
\usepackage{dsfont}
\usepackage{url}
\usepackage{verbatim}
\usepackage{graphicx}
\usepackage{cite}
\usepackage{booktabs}
\usepackage{multirow}
\usepackage{tabularx}
\usepackage{makecell}
\usepackage{ulem}

\newcount\Comments  
\usepackage{color}
\definecolor{darkgreen}{rgb}{0,0.5,0}
\definecolor{purple}{rgb}{1,0,1}
\definecolor{orange}{rgb}{1.0, 0.5, 0.0}
\newcommand{\kibitz}[2]{\ifnum\Comments=0\textcolor{#1}{#2}\fi}

\begin{document}
\title{MARVEL: Multi-Agent Reinforcement-Learning for Large-Scale Variable Speed Limits}  

\author{Yuhang Zhang$^{\dagger*}$, Marcos Quinones-Grueiro$^{\dagger}$, Zhiyao Zhang$^{\dagger}$, Yanbing Wang$^{\dagger}$,\\ William Barbour$^{\dagger}$, Gautam Biswas$^{\dagger}$, Daniel Work$^{\dagger}$
\\
$^{\dagger}$Institute for Software Integrated Systems, Vanderbilt University, 37212, Nashville, TN
\\
$^{*}$yuhang.zhang.1@vanderbilt.edu


}
\maketitle



\begin{abstract}

Variable Speed Limit (VSL) control acts as a promising highway traffic management strategy with worldwide deployment, which can enhance traffic safety by dynamically adjusting speed limits according to real-time traffic conditions. Most of the deployed VSL control algorithms so far are rule-based, lacking generalizability under varying and complex traffic scenarios. In this work, we propose MARVEL (\textbf{M}ulti-\textbf{A}gent \textbf{R}einforcement-learning for large-scale \textbf{V}ariable sp\textbf{E}ed \textbf{L}imits), a novel framework for large-scale VSL control on highway corridors with real-world deployment settings. MARVEL utilizes only sensing information observable in the real world as state input and learns through a reward structure that incorporates adaptability to traffic conditions, safety, and mobility, thereby enabling multi-agent coordination. With parameter sharing among all VSL agents, the proposed framework scales to cover corridors with many agents. The policies are trained in a microscopic traffic simulation environment, focusing on a short freeway stretch with 8 VSL agents spanning 7 miles. For testing, these policies are applied to a more extensive network with 34 VSL agents spanning 17 miles of I-24 near Nashville, TN, USA. MARVEL-based method improves traffic safety by 63.4\% compared to the no control scenario and enhances traffic mobility by 58.6\% compared to a state-of-the-practice algorithm that has been deployed on I-24. Besides, we conduct an explainability analysis to examine the decision-making process of the agents and explore the learned policy under different traffic conditions. Finally, we test the response of the policy learned from the simulation-based experiments with real-world data collected from I-24 and illustrate its deployment capability.
\end{abstract}

\begin{IEEEkeywords}
Multi-Agent Reinforcement Learning, Traffic Control, Variable Speed Limit
\end{IEEEkeywords}


\section{Introduction} \label{intro}
\IEEEPARstart{A}{ctive} traffic management (ATM) strategies hold the potential to alleviate traffic congestion and enhance road safety. Among the various ATM strategies, variable speed limit (VSL) control has been proposed to dynamically adjust speed limits in response to real-time traffic conditions. VSL control has been deployed around the world since 1960s, and its potential impact on road safety and travel mobility has been investigated both in simulation and operational deployments~\cite{lin2004exploring, chang2011its, lu2014review, khondaker2015variable}.

In practice, many deployed VSL methods  are rule-based (see Table~\ref{table:application}) given their relative simplicity in terms of implementation, minimal calibration requirements, and straightforward explainability properties. Most of these algorithms are based on activation conditions which are reactive by nature, resulting in increased latency when slowing down traffic or restoring it to normal conditions. Hence, many rule-based methods induce unnecessary travel time delays~\cite{allaby2007variable}. 



Learning-based methods for VSL control have recently attracted the attention from the traffic control community~\cite{walraven2016traffic, zhu2014accounting, wang2019new, zheng2023coordinated}. 
Because learning-based methods work directly on observed traffic data, rather than using the data to calibrate a model of traffic dynamics, they have the potential to improve  performance~\cite{ren2020data}. Moreover, while learning-based approaches may be computationally intensive to train, they are typically efficient to execute. Learning-based approaches such as deep reinforcement learning (DRL) also open the door for proactive control policies, since they aim to maximize the future cumulative rewards. 

The first works to design learning-based controllers for VSL include~\cite{7949069, wu2018differential}, which focus on the control of the speed on a single gantry (i.e., at a single location on the roadway). A more recent study~\cite{zheng2023coordinated} proposes a learning-based approach for VSL control at four gantries. However, there is a need for learning-based methods specifically designed to tackle large-scale VSL control that can be applied to long corridors with many more gantries. 

In pursuit of learning-based VSL control methods that are eventually deployable, we focus our work considering three requirements. First, our method should be scalable and computationally efficient to manage corridors with a large number of gantries. As illustrated in Table~\ref{table:application}, existing VSL implementations usually span considerable freeway lengths, necessitating coordinated control among numerous controllers. Second, the method should be able to generalize well to different levels of driver compliance rates, given that this may vary from location to location. Third, the method should only require commonly available sensor data to make control decisions, and those decisions should obey operational constraints (such as rules on maximum speed steps allowable on US roadways~\cite{MUTCD2009}).

\begin{table*}
\centering
\caption{Summary of selected VSL systems deployed around the world.}
\label{table:application}
\begin{tabularx}{\textwidth}{cccccccc}
\toprule
\textbf{Region} & \textbf{Location} & \textbf{Year} & \textbf{Length} & \textbf{Algorithm} & \textbf{Response Type} & \textbf{Functionality} & \textbf{Reference} \\ \midrule
\multirow{6}{*}{Europe}        & A2, Netherlands   & 1992 & 20 km     & rule-based & reactive      & incident/congestion/weather & \cite{van1994control,smulders1992control}\\
                               & M25, England  & 1995 & 18 mile   & rule-based & reactive      & incident/congestion         & \cite{uk2004m25}\\
                               & E18, Finland  & 1999 & 14 km    & rule-based & reactive      & weather           & \cite{rama1999effects} \\
                               & A9, Germany   & 2006 & 11.2 mile & rule-based & reactive      & incident/congestion/weather & \cite{bertini2006dynamics} \\
                               & A20, Netherlands  & 2011 & 4.2 km    & rule-based & reactive      & congestion/night            & \cite{hoogendoorn2013assessment} \\ \midrule
\multirow{6}{*}{\multirowcell{2}[0ex][c]{North\\ America}} & NJT, New Jersey & 1960s & 148 mile & rule-based & reactive & incident/construction/congestion/weather & \cite{robinson2000examples} \\
                               & I-40, New Maxico & 1989  & 3 mile  & rule-based & reactive & incident/congestion/weather & \cite{robinson2000examples}  \\
                               & I-4, Florida & 2006 & 10 mile & rule-based & reactive & incident/congestion/weather & \cite{elefteriadou2012variable} \\
                               & I-5, Washington & 2010 & 7 mile & rule-based & reactive & congestion & \cite{pu2020full} \\ 
                               & I-24, Tennessee & 2023 & 17 mile & rule-based & reactive & incident/construction/congestion & \cite{zhang2022quantifying} \\ 
                               & \textbf{I-24}, \textbf{Tennessee} & \textbf{2024 (expected)} & \textbf{17 mile} & \textbf{AI-based} & \textbf{proactive} & \textbf{incident/construction/congestion} & - \\ \bottomrule
\end{tabularx}
\end{table*}

Motivated by the three requirements established above, the main contributions of this paper are the following:
\begin{itemize}
    \item We propose MARVEL, a novel multi-agent reinforcement learning (MARL) framework for large-scale VSL control considering real-world constraints aimed for deployment. We demonstrate scalability by training with 8 VSL agents, and then applying the learned policy on a network considering 34 VSL agents. The algorithm only requires commonly available traffic sensor data.
    \item We design a novel multi-objective reward function balancing safety and mobility, such that the learned policy substantially enhances traffic mobility in microsimulation across varying traffic conditions and levels of driver compliance, while preserving safety improvements compared to a benchmark algorithm currently deployed on I-24.
    \item We demonstrate the potential deployment capability of the MARVEL-based method by testing the response of the learned policy using real data collected from I-24, which is the first time, to the best knowledge of the authors, that the predicted action of a learning-based VSL control system has been generated and evaluated using real-world traffic data.
    
\end{itemize}

Compared to our previous work~\cite{10207650}, which introduces a preliminary MARL framework for large-scale VSL systems, the extensions in this work are as established next. First, we extend the action space from a coarse three-action set to a finer resolution set required for actual deployments. Second, we introduce spatially sequential decision-making mechanism amongst the agents to improve decision making that is compliant with operational requirements between consecutive gantries. Third, we revise the state and reward design to improve the general performance. We evaluate our trained policy on a micro-simulation environment of the I-24 Smart Corridor, a segment of I-24 near Nashville, TN area. Finally, we test the response of the learned policy using real data collected from I-24 Smart Corridor.

The trained policy from the proposed MARVEL framework is planned to be deployed on I-24 in 2024 as shown in Table~\ref{table:application}. This deployment is expected to be a pilot AI application in a traffic management system and serves as a starting point to explore how AI will influence real-world traffic performance.

The remainder of the article is organized as follows: Section~\ref{lr} presents a literature review for VSL control based on VSL performance and different methodology categories. Section~\ref{marl} introduces preliminary concepts of MARL and one of the most popular algorithms MAPPO. The problem formulation and MARL framework design for VSL control are presented in Section~\ref{method}. In section~\ref{experiment}, we describe the training and testing scenarios, as well as the parameter settings for the algorithm. The results and the respective discussions are included in Section~\ref{results}. Finally, the paper is concluded in Section~\ref{conclusion} where future directions of work are also discussed.

\section{Literature Review}
\label{lr}
This section offers a literature review of VSL systems, focusing specifically on their safety and mobility effects on traffic, as well as on the methods used for algorithmic design, including rule-based and feedback control, optimization and predictive control, data-driven and learning-based control.

\subsection{VSL Performance}
VSL control systems have proven to be useful in improving road safety by smoothing travel speeds to prevent the abrupt braking of drivers~\cite{de2018safety}. In a micro-simulation study, the authors in ~\cite{abdel2006evaluation} compute the real-time crash likelihood for a section on the I-4 freeway in Orlando. Their findings reveal that VSL may enhance traffic safety in medium-to-high-speed conditions. A separate study proposes a proactive VSL control algorithm~\cite{yu2014optimal}. The authors also verify that safety may be improved through speed homogeneity in a micro-simulation setting. In~\cite{pu2020full}, the authors conduct a Bayesian before-after analysis for a VSL system implemented on the I-5 freeway in Seattle. They find that the total crash count decreases significantly after applying VSL. 

The impact of VSL control systems on mobility has also been demonstrated under certain conditions. For example, the simulation studies in~\cite{1657497,6851128,7006741} demonstrate that VSL systems can improve throughput by avoiding the capacity drop phenomenon. Still, further field experiment is needed to fully demonstrate that VSL systems could lead to mobility improvement~\cite{van1994control,elefteriadou2012variable}. Recognizing the difficulty of practically deploying model-free and model-based VSL control algorithms to improve mobility in the field, we focus our work on the design of methods for VSL systems that can improve road safety without significantly deteriorating mobility.

\subsection{Control Methods}
\subsubsection{Rule-based and Feedback Control}
Early research on VSL primarily involved the design of algorithms encoding reactive rule-based logic. Real-time decisions are based on traffic characteristics such as volume, speed, and occupancy~\cite{khondaker2015variable}. The primary objective of these systems is to enhance road safety by preventing abrupt breaking and smoothing travel speeds~\cite{pu2020full,khondaker2015variable}. The micro-simulation study in~\cite{abdel2006evaluation} showed that VSL systems may significantly improve road safety, especially in medium-to-high-speed conditions. Rule-based VSL controllers are predominantly adopted for most field applications because of the simple operations. They have demonstrated the benefit of safety improvement by reducing incidents and increasing homogeneous traffic flow~\cite{lu2014review}.

Feedback control-based VSL design has also attracted significant attention from the traffic control community. The essence of feedback control lies in its ability to adjust the current traffic state toward a desired reference value. The authors in~\cite{carlson2010optimal} propose the concept of mainstream traffic flow control aimed at improving mobility. In this setting, VSL systems operating upstream of the potential bottleneck regulate inflow rate by maintaining the highway density around its critical value. Therefore, capacity drop is avoided. For example, the authors in~\cite{carlson2011local} design a two-loop feedback cascade controller with the critical density of the bottleneck as the set point. The speed limit rate is set as the control input with the objective of maintaining bottleneck density around critical density by regulating the upstream inflow. The study in~\cite{7006741} evaluates the aforementioned feedback controller in a micro-simulation scenario and~\cite{6851128} extends the controller to a scenario with multiple bottlenecks.

\subsubsection{Optimization and Predictive Control}
Model-based VSL systems that utilize optimization techniques make decisions based on the estimated or predicted evolution of traffic dynamics. These dynamics are typically formalized through a macroscopic traffic model, which is formulated in terms of partial differential equations~\cite{lighthill1955kinematic,richards1956shock,aw2000resurrection}. Hegyi \textit{et al.}~\cite{hegyi2005model} present the integration of ramp-metering and VSL control as a model predictive control (MPC) problem, demonstrating that coordinated control enhances outflow. The same authors propose another MPC framework based on METANET~\cite{kotsialos2002traffic} model to involve safety constraints in work~\cite{hegyi2005optimal}. To overcome the deficiency of high computation load of METANET-based MPC, Han \textit{et al.}~\cite{han2017resolving} develop a more accurate discrete first-order traffic model that enables a fast MPC for VSL and illustrate its performance in a simulation study.

In addition to MPC-based VSL systems, other studies consider optimal control~\cite{carlson2010optimal, wang2020freeway} and genetic algorithms~\cite{yu2018optimal} for VSL design. Carlson \textit{et al.}~\cite{carlson2010optimal2} present the integration of ramp-metering and VSL control as an optimal control problem and show that traffic mobility can be improved substantially.  The results demonstrate a better performance both in reducing travel time and lowering computation time when compared with other MPC methods.
Yu and Abdel-Aty~\cite{yu2014optimal} propose a proactive VSL control by employing optimization technique to minimize the total crash risk. They find that the proposed approach can effectively reduce crash risk and enhance speed homogeneity in high and moderate compliance levels. While model-based VSL systems with optimization techniques could achieve the optimal solution theoretically, the execution of them in near real-time conditions is sometimes prohibitive, which makes them challenging to be deployed.

\subsubsection{Data-Driven and Learning-Based Control}
In recent years, the interest in developing learning-based algorithms for VSL control has grown steadily among researchers. As one of the most promising methods for sequential decision-making, reinforcement learning (RL) has been applied to solve the VSL control problems~\cite{kuvsic2020overview, wu2018differential, schmidt2015decentralised}. The authors in~\cite{7949069} propose a Q-learning-based algorithm to reduce travel time considering a simulation setting with a single VSL controller and the assumption of capacity drop. In~\cite{walraven2016traffic}, a similar VSL control problem is framed to reduce traffic congestion through multiple VSL controllers considering Q-learning as the algorithm for decision-making. However, they implement identical speed limits for all VSL controllers ignoring the potential coordination among them. The study~\cite{zhu2014accounting} considers a relatively large urban network as their testbed and verifies their proposed RL-based VSL control algorithm for mobility improvement and emission reduction. Notwithstanding, coordination among agents (VSL controllers) is not accounted for thus only local optimum performance is achievable. 

There are few MARL-based VSL studies in the literature. To the best of our knowledge, the authors in~\cite{wang2019new} are the first to formulate VSL into a MARL problem. They consider a vehicle-to-infrastructure (V2I) environment in which the connected vehicles will guarantee driving at a specified speed. However, such an assumption makes deployment difficult because of the current lack of a viable V2I setting in which to test. Moreover, the scalability of the algorithm needs to be further investigated as there are only three control segments in the simulation case study considered. In~\cite{zheng2023coordinated}, the authors apply the MADDPG algorithm to four VSL controllers to improve traffic mobility with a consideration of a capacity drop.
With a similar idea to the feedback controllers, they design a reward function to control the bottleneck density to avoid the capacity drop. However, avoiding capacity drop needs a high driver compliance level, which could be hard to achieve for areas such as the south of the United States where the proposed approach will be deployed. Besides, it remains unknown how the algorithm performs in terms of generalizability and scalability.

\section{Preliminaries}
\label{marl}
In this section, we first introduce the notation and the general idea of RL. We review the preliminary concepts of MARL and describe one of the state-of-the-art algorithms in MARL called MAPPO. 

\subsection{RL and MARL}
Reinforcement learning focuses on finding a solution to sequential decision-making problems where an agent is expected to learn a policy by interacting with the environment through trial and error. Formally, RL is formulated as a Markov Decision Process (MDP), which can be defined by the tuple $\langle \mathcal{S}, \mathcal{A}, \mathcal{R}, P, \gamma \rangle$, where $\mathcal{S}$ denotes the state space, $\mathcal{A}$ denotes the action space, $\mathcal{R}: \mathcal{S \times A \times S} \rightarrow \mathbb{R}$ denotes a scalar reward, $P: \mathcal{S \times A \times S} \rightarrow [0, 1]$ denotes a probability transition function and $\gamma \in [0, 1]$ denotes the discount factor. The goal of RL algorithms is to maximize the cumulative discounted reward: 
\begin{align}
    J(\theta)=\mathbb{E}_{s_t, a_t \sim \pi_{\theta}(a_t|s_t)}\left[\sum_{t=0}^T\gamma^tr_t\right]
\end{align}
where $s_t, a_t, r_t$ are the state, action, and reward at time step $t$, respectively. In the context of learning-based models, neural networks are generally used to learn a policy $\pi_{\theta}$ (a mapping between the state and actions). Here $\theta$ denotes the neural network parameters. 

MARL extends the standard RL paradigm to environments with multiple agents. Formally, the MARL problem can be modeled as a Markov Game, defined as a tuple $\langle\{\mathcal{S}^i\}_{i\in \{1,\dots, n\}}, \{\mathcal{A}^i\}_{i\in \{1,\dots, n\}}, \{\mathcal{R}^i\}_{i\in \{1,\dots, n\}}, P, n, \gamma\rangle$, where $\mathcal{S}^i$: denotes the local state space for agent $i$, $\mathcal{A}^i$: denotes the action space for agent $i$, $\mathcal{R}^i$: $\{\mathcal{S}^i\}_{i\in \{1,\dots, n\}} \times \{\mathcal{A}^i\}_{i\in \{1,\dots, n\}} \times \{\mathcal{S}^i\}_{i\in \{1,\dots, n\}}\rightarrow \mathbb{R}$ denotes the reward for agent $i$, $P$: $\{\mathcal{S}^i\}_{i\in \{1,\dots, n\}} \times \{\mathcal{A}^i\}_{i\in \{1,\dots, n\}} \times \{\mathcal{S}^i\}_{i\in \{1,\dots, n\}}\rightarrow [0, 1]$ denotes the transition probability of the environment from a given state $s$ to the next one $s'$, and $n$ denotes the total number of agents in the environment.

Similar to single-agent RL, the goal of MARL for each agent is to learn a policy that maximizes its own cumulative discounted reward: 
\begin{align}
    J^i(\theta_1, \dots, \theta_n)=\mathbb{E}_{s_t, A_t}\left[\sum_{t=0}^T \gamma^t r_t^i \right],
\end{align}
where $s_t$ denotes the global state at time step $t$ and $A_t=(a_t^1, \dots, a_t^n)$ denotes the joint action of all agents at time step $t$. It is well known that MARL is a more challenging problem since multiple agents are simultaneously interacting with the environment and making decisions that affect each other's rewards, which will raise issues such as non-stationarity and the credit assignment problem. Therefore, in the context of problems that require cooperation among the agents to find a solution, it is hard for them to learn to cooperate with each other to achieve a common goal while maximizing their individual rewards at the same time.

MARL algorithms can be broadly classified into two categories: centralized and decentralized learning. In fully centralized algorithms, there is a central agent that acquires information about all states and actions and outputs a joint action. This approach reduces the MARL problem to a single agent but with a much larger state and action space. This method is generally inefficient since learning a central controller with large action or state space costs time. Therefore, it has been shown that a fully centralized MARL could fail even for simple tasks~\cite{sunehag2017value}. In contrast, fully decentralized algorithms allow each agent to make decisions based only on its local observations. However, naive decentralized MARL methods are often unsuccessful because of the non-stationarity problem~\cite{sunehag2017value}. Recently, centralized training and decentralized execution (CTDE) framework has gained increasing attention in MARL research~\cite{foerster2016learning, lowe2017multi, rashid2020monotonic}. This framework involves training a centralized policy that observes the states and actions of all agents during training, while each agent executes a decentralized policy based on its local observations during deployment. The CTDE framework can improve the stability and scalability of MARL algorithms, especially in large-scale and complex environments, as it allows the agents to benefit from the centralized information during training while maintaining the flexibility and efficiency of decentralized execution.

\subsection{Policy Optimization Methods}
Policy optimization methods in standard RL focus on learning an explicit policy $\pi_{\theta}(a_t|s_t)$ by optimizing the objective function $J(\theta)$. While it is true that most policy optimization methods suffer from sample inefficiency, they usually demonstrate stable learning with monotonic performance improvement~\cite{schulman2015trust}. In general, policy gradient for learning neural networks has the following form: 
\begin{align}
    \nabla J(\theta)=\mathbb{E}_{s_t, a_t \sim \pi_{\theta}(a_t|s_t)}\left[\sum_{t=0}^T \nabla_{\theta} \log \pi_{\theta}(a_t|s_t)\Phi_t \right]
\end{align}
where $\Phi_t$ is usually represented by an estimated advantage function in order to reduce variance. 

This work applies a policy optimization method called MAPPO~\cite{yu2022the}, which is a variant of PPO~\cite{schulman2017proximal} in a multi-agent setting. PPO is a widely used policy optimization method that applies a clipped surrogate objective to prevent large policy updates trying to guarantee stable learning. PPO utilizes an actor-critic architecture where the actor learns a policy while the critic learns a state value function $V_{\phi}(s_t)$ that estimates the cumulative reward for a given state. 

MAPPO extends PPO to the multi-agent setting by including the following features:
\begin{itemize}
    \item Adoption of CTDE where each agent has its own actor network with local observations as input and individual actions as output while a centralized value function to estimate the global state-value function is implemented to facilitate coordination.
    \item PopArt~\cite{hessel2019multi} is implemented to gain a more stable value function learning.
    \item A small value for the mini-batch parameter is suggested in order to tackle the non-stationarity issue. 
\end{itemize}
For more details, please refer to~\cite{yu2022the}.
\begin{figure}
    \centering
    \includegraphics[width=\columnwidth]{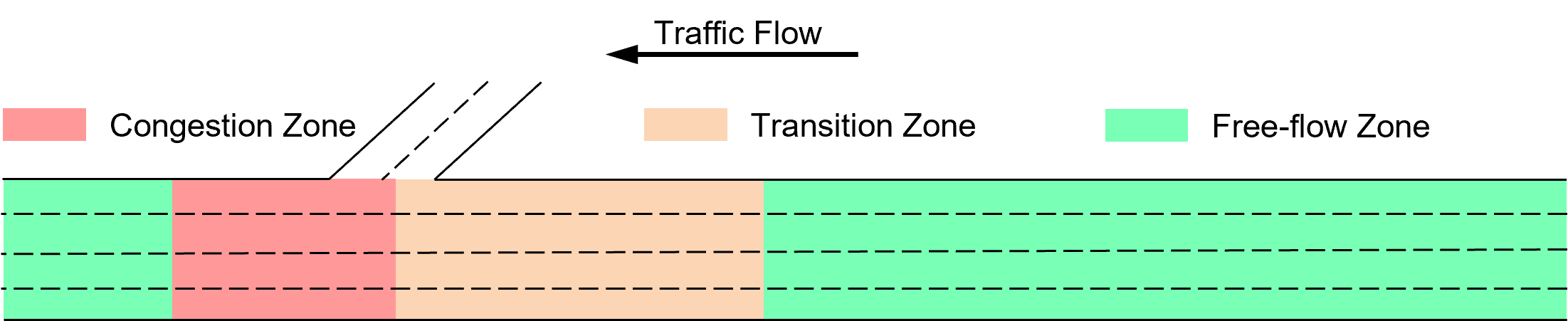}
    \caption{The three traffic zones for VSL design.}
    \label{design}
\end{figure}

\section{Methodology}
\label{method}
In this section, we formulate the large-scale VSL control problem over multiple gantries as a cooperative MARL problem. We aim to develop a collaborative approach involving multiple agents that targets the subsequent objectives:
\begin{itemize}
    \item Adaptability: The recommended speed limits should closely reflect actual traffic speeds to enhance drivers' comfort levels with these limits when the traffic is in congestion state.
    \item Safety: By reducing spatial speed variation across areas, the speed limits defined should enhance road safety.
    \item Mobility: While ensuring safety and adaptability, the speed limits should not substantially compromise traffic flow or mobility.
\end{itemize}

The aforementioned three objectives are designed to guide the driver's behavior across the three traffic zones depicted in Figure~\ref{design}. Each agent must adapt to the surrounding traffic conditions and make decisions to inform appropriate speed limits to the drivers on the road. For example, when an agent is located in a congestion zone, it should post the minimum speed limit to reflect the congestion state. On the contrary, if the traffic condition is or could be free flow, the same agent is supposed to post the maximum speed limit. Agents located in the transition zone should coordinate to post a smooth slow-down speed profile to reduce speed variation and prevent abrupt braking. To achieve these goals, we present next a set of design considerations constituting the MARVEL framework.

\subsection{Temporal and Spatial Sequential Decision Making} \label{sdm}
RL is renowned for addressing problems of sequential decision-making, often termed as ``temporal sequential decision-making''. In this research, we augment the MARL framework to incorporate real-world step-down constraint, which enforces the maximum speed limit difference between two consecutive gantries for the slow-down profile. Specifically, we recast the problem to encompass both temporal and spatial dimensions of sequential decision-making. This means that while every agent selects an action at each time step $t$—representing the temporal dimension—we define agents to make decisions in a sequence at a fixed time step $t$, starting from the most downstream agent. Through this setup, we facilitate communication, allowing the preceding agent to share information with the following agent. The detailed definition of agent, state space, and action space in this context are as follows:

\subsubsection{Agent}
Each agent will act on a VSL controller deployed to a highway gantry at a specific fixed location. 
This setting is more realistic as each agent has the flexibility to make its own decision according to the location-based traffic condition. Meanwhile, as all agents have the same type of state space, action space, and objective, we consider this problem as a homogeneous problem where all agents can share the same parameters.  

\subsubsection{State Space}
The goal of state space design is to inform each agent about the traffic conditions upstream and downstream of the gantry’s location. In our previous work~\cite{10207650}, the state space design involves speed, volume, and occupancy extracted from commonly available Radar Detection Systems (RDS) on highways. However, based on feature selection studies, we have found that volume tends to be uninformative and might potentially mislead the agent. 
Therefore, volume is excluded as part of the state space of each agent in this study. To enhance the cooperation among agents, we incorporate the preceding agent's selected action into the following agent's observation. Concretely, the state for agent $i$ can be defined as a tuple of five elements: $\langle a_t^{i-1}, \nu_t^i, o_t^i, \nu_t^{i+1}, o_t^{i+1} \rangle$. Here $a_t^{i-1}$ is the action selected by the preceding (downstream) agent, which is set as the maximum speed limit for the most downstream agent, while $\nu_t^i, o_t^i, \nu_t^{i+1}, o_t^{i+1}$ denote the average speed and occupancy recorded by the RDS units closest to the location of the agent and the upstream agent, respectively.

\subsubsection{Action Space}
Building upon our previous research, we define the action space for each agent as a discrete set comprising values of 30, 40, 50, 60, and 70 mile/hr. These values represent the potential speed limits to be displayed. It's worth noting that the granularity of this action space is sufficiently detailed for practical field deployment.

\subsection{Reward Design} 
Designing a reward function for cooperative agents with multiple objectives is not trivial, especially when the objectives are conflicting. While multi-objective reinforcement learning (MORL)~\cite{hayes2022practical} approaches, which leverage a reward vector in place of a scalar, are primarily designed to tackle such complex problems, the overall performance of these methods remains inconclusive due to the scarcity of successful real-world applications in the literature.

Here we apply a linear combination of reward terms to optimize different objectives. While our prior research~\cite{10207650} utilized both local and global reward categories within a coarser action space, the present study focuses exclusively on local terms. We have observed that this configuration yields superior performance. The formal definition of each reward term is presented below:

\subsubsection{Adaptability}
The adaptability term is used to penalize an agent posting high-speed limits when the traffic is in congestion and is used to help the agent to identify the congestion state. It can be described as:
\begin{align}
    r_{1t}^i=\begin{cases}
        -10 & \text{if $\nu_t^i \leq 35$ and $a_t^i \neq 30$} \\
        0 & \text{otherwise}
    \end{cases}
\end{align}
where $a_t^i$ denotes the selected action of agent $i$ at timestep $t$, and $\nu_t^i$ denotes the real average traffic speed recorded by the RDS unit of agent $i$.

\subsubsection{Safety (step-down penalty)}
In our prior research~\cite{10207650}, we incorporated both the``safety" and ``step-down violation penalty" terms to encourage cooperative behavior among agents, aiming to reduce spatial speed variation. However, we found that these two terms introduce redundancy, adding unnecessary complexity to the overarching reward function design. Consequently, we have refined our approach to create a more streamlined reward design, emphasizing agents' adherence to step-down rules within transition areas:
\begin{align}
    r_{2t}^i=
    \begin{cases}
        0 &
        \begin{aligned}
            &\text{if (agent $i$ is at most downstream)} \\
            &\text{or ($a_t^{i-1}=30$ and $a_t^i \in \{30,40\}$)}
        \end{aligned} \\
        \\
        +2 &
        \begin{aligned}
            &\text{if ($a_t^{i-1}=40$ and $a_t^i=50$)} \\
            &\text{or ($a_t^{i-1}=50$ and $a_t^i=60$)} \\
            &\text{or ($a_t^{i-1} \in \{60,70\}$ and $a_t^i=70$)}
        \end{aligned} \\
        \\
        -2 \times\frac{a_t^i-a_t^{i-1}}{a_{\text{diff}}} & \text{if $a_t^i > a_t^{i-1}+a_{\text{diff}}$}
    \end{cases}
\end{align}
where $a_t^{i-1}$ denotes the selected action of the preceding (downstream) agent at time-step $t$, and $a_{\text{diff}}$ denotes the step-down value, which is commonly set as $10$ mile/hr in real-world. The most downstream agent receives neither reward nor penalty due to the absence of a preceding downstream agent. If an agent has a preceding agent in the congestion area posting the minimum speed limit, we neither reward nor penalize it, regardless of whether it selects the minimum speed due to propagated congestion or chooses a higher limit in line with the step-down rule. This design minimizes conflicts between reward terms and reduces the potential for unintended behaviors. An agent earns a positive reward for adhering to the step-down rule and a penalty for violations.

\subsubsection{Mobility}
The mobility term is used to encourage the agent to post a high-speed limit when the traffic condition is free-flow. It can be described as:
\begin{align}
    \nu_{\text{clip}}^i &= \begin{cases}
        \nu_t^i & \text{if $\nu_t^i \leq \nu_{\text{max}}^i$} \\
        \nu_{\text{max}}^i & \text{otherwise}
    \end{cases}\\
    r_{3t}^i&=\frac{\exp{(\frac{\nu_{\text{clip}}^i}{\nu_{\text{max}}^i})}-\exp{(0)}}{\exp{(1)}-\exp{(0)}}
\end{align}
where $\nu_{\text{max}}^i$ is a normalizer set to 70 in our experiments. Here we apply a nonlinear exponential function to enhance the reward sensitivity for the previously defined discrete action space.

Finally, the reward function for agent $i$ at time-step $t$ is the following:
\begin{align}
    r_t^i&=w_1 r_{1t}^i + w_2 r_{2t}^i + w_3 r_{3t}^i
\end{align}
where in this study, we set $w_1=0.2, w_2=0.3, w_3=0.5$. These particular values were selected via hyperparameter search and found to yield improved performance. As it can be seen, by integrating the presented three reward terms, the agents are encouraged to first identify the congestion state and post the minimum speed limit, then generate a slow-down speed profile for upstream drivers to reduce speed variation, and eventually improve the mobility as much as possible.


\begin{figure*}
    \centering
    \includegraphics[width=\textwidth]{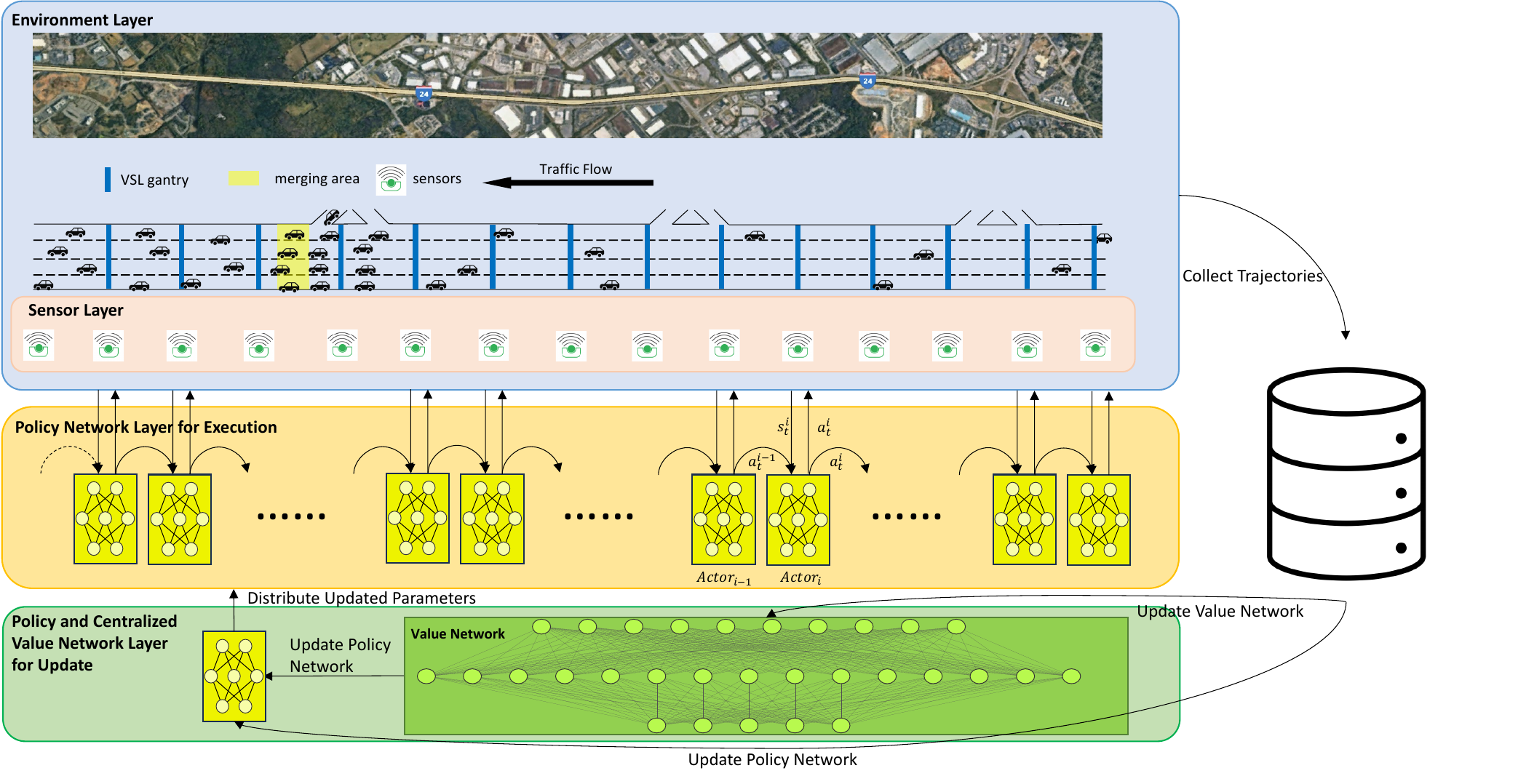}
    \caption{The VSL control framework based on MAPPO algorithm.}
    \label{framework}
\end{figure*}

\subsection{Invalid Action Masking}
As mentioned before, the implementation of VSL in the real world is often associated with rule-based constraints such as the step-down method described before. Although we explicitly involve a large penalization for the states where the step-down rule is violated, this still does not guarantee that there will never be. We therefore consider defining a safety layer after the output of the policy network to prevent step-down violations during the testing scenarios. The rationale of this safety layer is based on the invalid action masking method~\cite{huang2020closer}, which serves to “mask out” invalid actions and just sample from the set of valid ones. In this study, we define an invalid action set according to the following equation:
\begin{align}
    I=\{a | a > a_t^{i-1} + a_{\text{diff}}\}
\end{align}
where $a_t^{i-1}$ is the action selected by the preceding (downstream) agent. Note that for the first agent in the corridor, we assume that the selected action from the preceding one is always $70$, and therefore there is no invalid action in this case.

\begin{algorithm}
\caption{MARVEL-MAPPO}
\begin{algorithmic}[1]

\State Initialize policy network $\pi_{\theta}$, centralized value network $V_{\phi}$, number of PPO epochs $E$ and number of mini\_batch $K$
\label{pseudocode}
\While{training\_step $<$ training\_step\_max}   
    \State Set data buffer $D = \{\}$
    \For{t = 1 to batch\_size}
        \For{agent i = 1 to n}
            \State \parbox[t]{0.7\columnwidth}{Receive $s_t^i=\langle a_t^{i-1}, \nu_t^i, o_t^i, \nu_t^{i+1}, o_t^{i+1} \rangle$ from preceding agent and the sensor layer}
            \State \parbox[t]{0.7\columnwidth}{Generate $a_t^i$ from the policy network $\pi_{\theta}(s_t^i)$}
        \EndFor
        \State \parbox[t]{0.8\columnwidth}{Execute actions $a_t=\{a_t^1,\dots,a_t^n\}$, observe $r_t=\{r_t^1,\dots,r_t^n\}$ and $s_{t+1}=\{s_{t+1}^1,\dots,s_{t+1}^n\}$}
        \State $D=D\cup(s_t,a_t,r_t,s_{t+1})$
    \EndFor
        
    \For{ppo\_epoch e = 1 to E}
        \For{mini\_batch k = 1 to K}
            \State $b \leftarrow$ random mini-batch from $D$
            \State Adam update $\theta$ on $L(\theta)$ with data $b$ and $V_{\phi}$
            \State Adam update $\phi$ on $L(\phi)$ with data $b$
        \EndFor
    \EndFor
\EndWhile
\end{algorithmic}
\end{algorithm}

\subsection{VSL Control Framework for MAPPO}
Figure~\ref{framework} displays the VSL control framework based on the MAPPO algorithm but it can be easily generalized to other MARL algorithms. The general pipeline has three layers and one data buffer.

The environment layer includes all simulation-related components in the training scenario, which is discussed in the next section. Nested within the environment layer is the traffic sensor layer, responsible for providing agents with relevant traffic information, as discussed in Subsection~\ref{sdm}. 

The execution layer contains the policy (actor) network of each agent. At time-step $t$. Every agent makes a decision sequentially starting from the most downstream one. For example, after agent $i-1$ selected action $a_t^{i-1}$, the agent $i$ will get the combination of the corresponding traffic characteristics and $a_t^{i-1}$ as part of its state space to make its own decision. The selected speed limits will not be implemented until all agents finish their decision-making at a fixed time-step. 

The update layer is used to update the parameters of both the policy and value network. After collecting a certain amount of trajectories into the data buffer, the framework will use the collected trajectories to update the value network. The policy network will be updated based on both data trajectories and the value network. The updated policy network will be distributed to every agent to interact with the environment layer because of the assumption of homogeneity. 

 The algorithm begins by initializing the policy network $\pi_{\theta}$, the centralized value network $V_{\phi}$, and setting parameters such as the number of epochs $E$ and the number of mini-batches $K$. The training loop iterates until reaching the maximum number of training steps which are predefined in advance but can be modified as necessary. Within each training step, the algorithm collects data by receiving observations and actions from multiple agents, generates actions based on the policy network, executes these actions in the environment, and stores the resulting experience in a buffer $D$. Then,  we perform MAPPO updates over several epochs and mini-batches, optimizing the parameters $\theta$ and $\phi$ using the collected data and the centralized value network. The algorithm aims to train a set of cooperative agents to learn a policy that maximizes the expected cumulative reward in a value-sensitive manner. The detailed pseudo-code can be referred to Algorithm~\ref{pseudocode} and the details of MAPPO algorithm can be referred to~\cite{yu2022the}.

\section{Experiments}
\label{experiment}

In this section, we discuss the experimental design for both training and testing scenarios. The implementation details of the algorithm will be listed at the end of this section. We use TransModeler~\cite{balakrishna2009large,yang1996microscopic}, a commercial microscopic traffic simulator allowing customization of speed limits from Python through GISDK API, for both training and testing scenarios.

\subsection{Training Scenario} 
The training scenario considers a 7-mile long corridor segment with four lanes on I-24 westbound, Nashville. Three ramps present along this stretch are included in the experiment network. Most existing RL-based VSL studies only focus on analyzing the performance across a few number of VSL controllers thus lacking the ability to understand the full potential of VSL systems. In this work, we set up VSL at 0.5-mile intervals for training purposes, which results in 15 VSL controllers. Based on a trial and error approach, we deploy 8 VSL controllers upstream of the on-ramp merging location to serve as agents since they can manage diverse traffic conditions and develop a cooperative control policy, while also minimizing computational requirements. The rest of the 7 VSL controllers are set as the maximum speed limit during training. A common visibility parameter representing the effective control area in the simulation is applied to every VSL controller (agent). One RDS unit aggregating traffic data over 60 seconds is collocated with each VSL's gantry to capture traffic conditions. The simulation starts at 7:50 AM and ends at 10:00 AM with the first 10 minutes as a warm-up period for the simulation.

To introduce recurring congestion induced by on-ramp traffic weaving behavior, we set two lanes for the first downstream on-ramp with a flow around 1000 veh/lane/hr, which is fixed over the full simulation period. The mainstream inflow is set high enough as 1850 veh/lane/hr for the first hour to generate recurring congestion. The mainstream inflow then reduces to half for the second hour to allow clearance of the congestion. This setting mimics a full cycle of recurring congestion from the formation to the full clearance and therefore is selected as a training scenario. Moreover, we consider a compliance rate of 5\% during training as we believe the real-world compliance rate is relatively low in certain areas of the U.S., especially for the deployment site that we are considering. The simulation network is shown in Figure~\ref{framework}.

\subsection{Testing Scenario}
\begin{figure*}
    \centering
    \includegraphics[width=\textwidth]{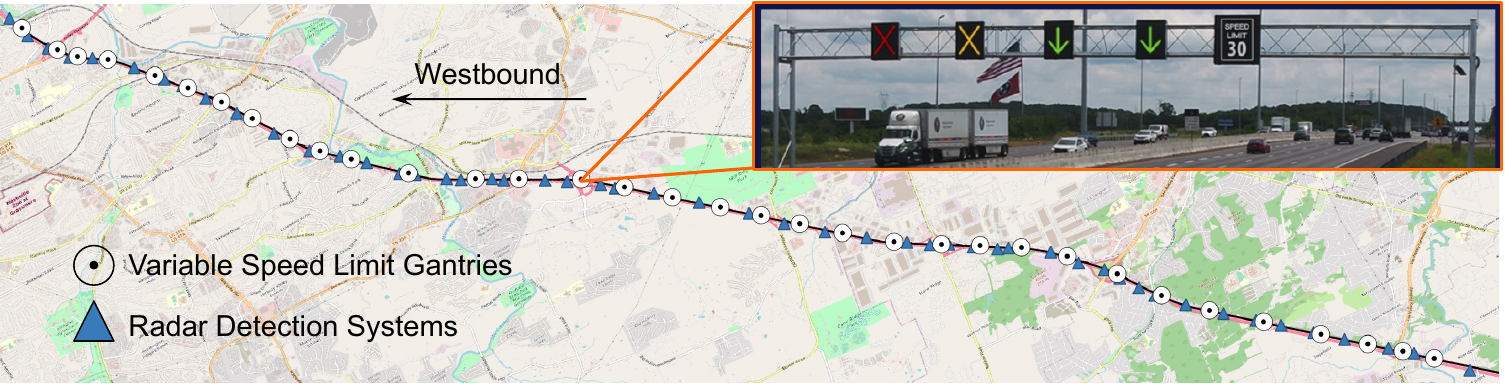}
    \caption{The I-24 Smart Corridor segment with VSL control.}
    \label{smartcorridor}
\end{figure*}
The testing scenario is used to evaluate the generalizability and scalability of the learned policy. By customizing the network geometry, the mainstream traffic demand, and the compliance rate, we can test the robustness of the learned policy across different traffic scenarios. Similarly, we can test the scalability of the learned policy by increasing the number of implemented agents. 

Specifically, we consider a longer segment of I-24 as our testing scenario, which has been selected as Tennessee's first testbed for an Integrated Corridor Management project. The selected segment, which is so-called I-24 Smart Corridor, spans around 28 miles long between Nashville and Murfreesboro. The physical infrastructure for VSL control has already been installed on-site as shown in Figure~\ref{smartcorridor}, which covers 17 miles from milemarker 53 to 70. The gantries used to display VSL signals are nearly uniformly distributed with an average distance of 0.5 miles between each other. The RDS units are installed to capture traffic characteristics with locations as shown in Figure~\ref{smartcorridor}.

For simulation purposes, this 17-mile segment is replicated in TransModeler, with a specific focus on westbound traffic, encompassing a total of 34 VSL gantries. To balance simplicity with fidelity, a single sensor is assigned to each gantry and placed at a stochastic distance ranging from 0 to 0.2 miles from the gantry. We will discuss how to map multiple sensors into one gantry in the next section, where we will show the behaviors of the learned policy based on the real-time traffic data from all RDS units as shown in Figure~\ref{smartcorridor}.

To have a comprehensive evaluation of the trained policy, we consider three testing scenarios as follows. The testing \textbf{\textit{Scenario A}} incorporates a multi-bottleneck traffic configuration, containing two recurrent congestion induced by on-ramp traffic merging. This design choice facilitates an assessment of the policy's generalizability across typical multi-bottleneck situations, which are frequently observed on I-24 during peak hours. To enhance the simulation's fidelity, we employ time-varying inflow rates for both the mainstream and ramp traffic. The compliance rate for the simulated scenario is set at 5\%. The testing \textbf{\textit{Scenario B}} and \textbf{\textit{C}} are designed to test the learned policy's generalizability with respect to compliance rate, in which we only set one ramp merging zone and compliance rate as 50\% and 100\% respectively. The traffic demands are set lower than Scenario A to make sure the congestion does not propagate out of the network boundary. All three scenarios include a three-hour simulation.

In summary, the testing scenario introduces considerably more agents than the training one, thereby enhancing the complexity and rigor of the evaluation. Additionally, the variability in network geometry, significantly different compliance rates, and the setting of multiple bottlenecks as well as the stochasticity of the micro-simulation present further challenges, providing a robust test for the learned policy's adaptability and effectiveness.


\subsection{Benchmark Algorithms and Performance Metrics}
In evaluating the effectiveness of VSL control algorithms, we have selected the following scenarios for benchmarking purposes:
\begin{itemize}
    \item \textbf{No Control}: The baseline scenario where no VSL control is present on the corridor. The default speed limit for each gantry is 70 miles/hr.
    \item \textbf{Speed-Matching (currently deployed on I-24)}: The scenario with a rule-based VSL control algorithm, which is the most popular category of algorithms used for VSL deployments (See Table~\ref{table:application}). The rule-based speed-matching algorithm is designed to display a speed limit value that is as close as possible to the slowest speed measured on the highway within a certain distance limit to the respective gantry. Specifically, the VSL controller will be triggered whenever the real traffic characteristics (volume, speed, occupancy) violate some predefined thresholds for a certain amount of time, and the speed limit will be rounded to the nearest multiple of 10 mph. To satisfy the step-down rule and to reduce the speed variation, the upstream gantries will have a sequential increment of 10 mph. The available speed limit values have a lower bound of 30 mph and an upper bound of 70 mph. This algorithm has been implemented on the I-24 Smart Corridor and demonstrated its effectiveness in reducing the car crash rate compared to no VSL control. The thresholds used for this research have been carefully searched to achieve optimal performance.
    \item \textbf{MARVEL-IPPO}: MARVEL framework applied to independent proximal policy optimization algorithm, in which each agent has its own policy and critic network and behaves independently in the environment. The parameters are shared by all agents.
    \item \textbf{MARVEL-MAPPO-IAM}: The same with MARVEL-MAPPO but with invalid action masking (IAM) during testing. This benchmark can be used to evaluate whether IAM will introduce a negative effect on the general traffic performance while guaranteeing the step-down rules.
    
\end{itemize}



The mobility performance is evaluated using the queue length, which is defined as the total length of the congestion queue where the average traffic speed is lower than 35 miles/hr. The safety performance is evaluated using the normalized coefficient of variation in speed ($\overline{\text{CVS}}$)~\cite{10207650}, which is defined as follows:

\begin{align}
    \text{CVS}_t^i = \begin{cases}
        \frac{\sigma_t^i}{\bar{v}_{t}^i }& \text{if } v_{t}^i \leq \bar{v}_{t}^i\\
       0 & \text{otherwise}
    \end{cases}
\end{align}
\begin{align}
    \overline{\text{CVS}}=
        \frac{\sum_{t=1}^T\sum_{i=1}^n\left(\text{CVS}_t^i\right)_{>\alpha}}{\sum_{t=1}^T\sum_{i=1}^n\left(\mathds{1}_{\text{CVS}_t^i>\alpha}\right)}
\end{align}
\noindent where $\text{CVS}_t^i$ represents the coefficient of variation in speed (CVS) for agent $i$ at time $t$, in which $\sigma_t^i$ and $\bar{v}_{t}^i$ are calculated based on the collocated RDS speed ($v_t^i$) and the closest upstream RDS ($v_t^{i+1}$) speed. To capture the large variation, we apply a truncation function $(x)_{>\alpha}$ to CVS, which returns the value $x$ if $x>\alpha$, and 0 otherwise. $\mathds{1}_{x>\alpha}$ returns the value of $1$ if $x>\alpha$ and $0$ otherwise. Here we select $\alpha$ as $0.1$.


To have a thorough understanding of how MARVEL-based methods perform from the perspective of design objectives, we introduce two other metrics, i.e., adaption violation number and step-down violation number. The adaption violation number is defined as the number of times that an agent does not post the minimum speed limit when the area under its control is in a congested state (i.e., $\nu_t^i<=35$); and the step-down violation number counts how many times the agents are violating the step-down rule in a single simulation run. By involving these two metrics, we can evaluate the performance of MARVEL-based methods for each reward term as well as the general traffic performance.

\subsection{Implementation Details}
Table~\ref{table:hyperparameter} shows the hyper-parameter settings for the MAPPO and IPPO algorithms in the training scenario. Since we consider this problem as homogeneous from the agents' standpoint, we share the parameters among all agents for both the actor and critic networks. All experiments are performed on a local machine with a 16-core CPU, 16 GB RAM, and an NVIDIA GeForce RTX 3060 GPU. The default car-following model and parameters in TransModeler are selected for both training and testing scenarios~\cite{ahmed1999modeling}, with a purpose of evaluating all methods based on a state-of-the-art commercial microsimulator.


 \begin{table}[h]
\centering
\caption{Parameter settings for training scenario}
\label{table:hyperparameter}
\begin{tabularx}{0.5\columnwidth}{Xc}
\toprule
Parameters   & Values  \\ 
\midrule
Episode length: &120 \\
Training batch size: &120 \\     
Number of mini-batch: &1 \\        
Number of hidden layer: &2 \\     
Hidden layer size: &64 \\  
Actor learning rate: &7e-4 \\
Critic learning rate: &5e-4 \\
Number of PPO epochs: &15 \\
Entropy coefficient: &0.05 \\  
Value loss coefficient: &1 \\
Discount factor: &0.99 \\
\bottomrule
\end{tabularx}
\end{table}

\begin{figure*}[h]
    \centering
    \includegraphics[width=\textwidth]{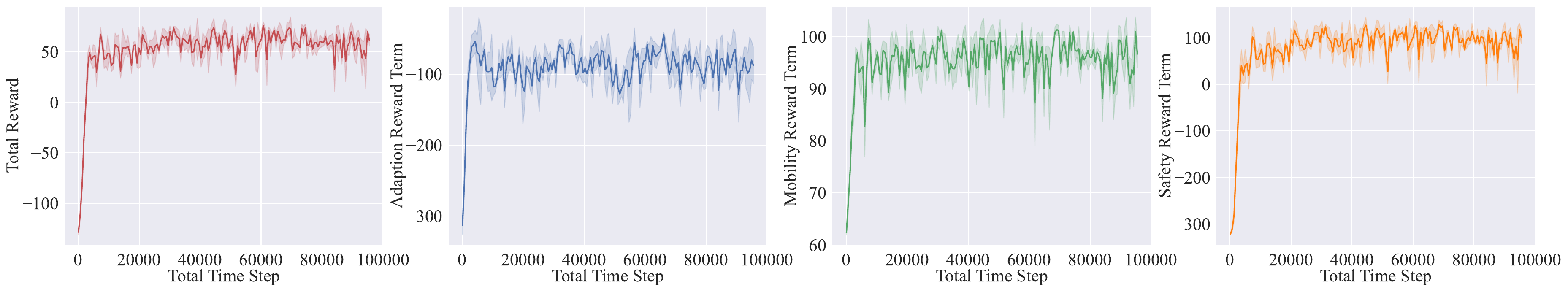}
    \caption{The learning curve of the total reward, adaption reward term, mobility reward term and safety reward term.}
    \label{reward_curve}
\end{figure*}

\section{Results and discussion}
\label{results}

In this section, we present the training performance of MARVEL-MAPPO proposed in Section~\ref{method} and evaluate its performance under testing scenarios along with other benchmark algorithms. In addition, we conduct explainability analysis for agents' decision-making. Finally, we present MARVEL-MAPPO-IAM's behavior using empirical real RDS data.

\subsection{Training Performance}
One single training takes about 12.5 hours of wall-clock time on the aforementioned PC to finish. The training curve with three random seeds is presented in Figure~\ref{reward_curve}. On the one hand, we can observe that all agents are learning efficiently during the first 10,000 time steps where all reward terms increase to a certain level. On the other hand, all reward terms fluctuate thereafter, which can be attributed to the inherent complexities in learning coordination and the contradicting nature of the safety (step-down rule) and mobility objectives. Despite the oscillations, an overall increasing trend in the safety reward term's learning pattern can be appreciated. This trend means that the agents gradually improve their ability to balance safety and mobility over time, demonstrating the effectiveness of our designed reward function. These insights from the training curve reinforce the validity of our approach and highlight its potential for addressing more complex, real-world scenarios where adaptation, mobility, and safety are all crucial.

\begin{figure*}[h]
    \centering
    \includegraphics[width=\textwidth]{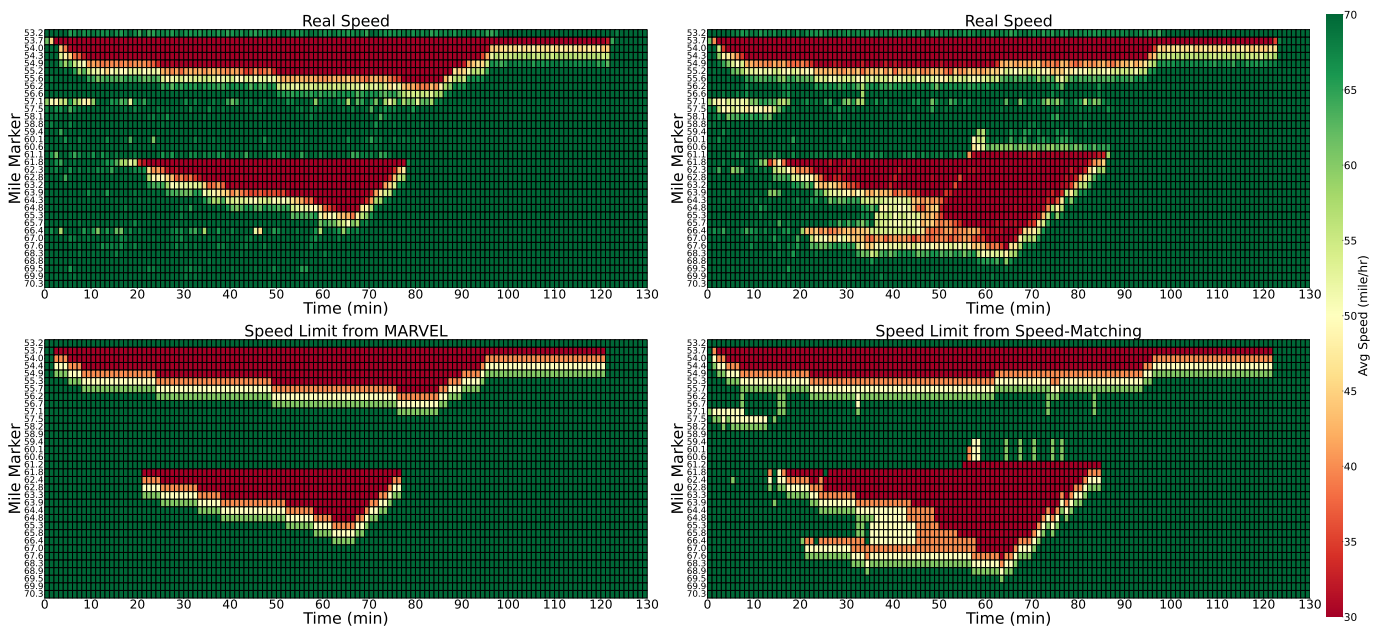}
    \caption{The time-space diagram of Scenario A. The left column shows the scenario under MARVEL-MAPPO control and the right column shows the scenario under speed-matching control. (Upper row: the average traffic speed from RDS units. Lower row: the control outputs from the algorithms. \textit{Note: the figure is trimmed in time to show the congestion better.})}
    \label{fig:scenario_a}
\end{figure*}
\subsection{Testing Performance: Qualitative Analysis}

\begin{figure*}
    \centering
    \subfloat[Scenario B]{
        \includegraphics[width=\textwidth]{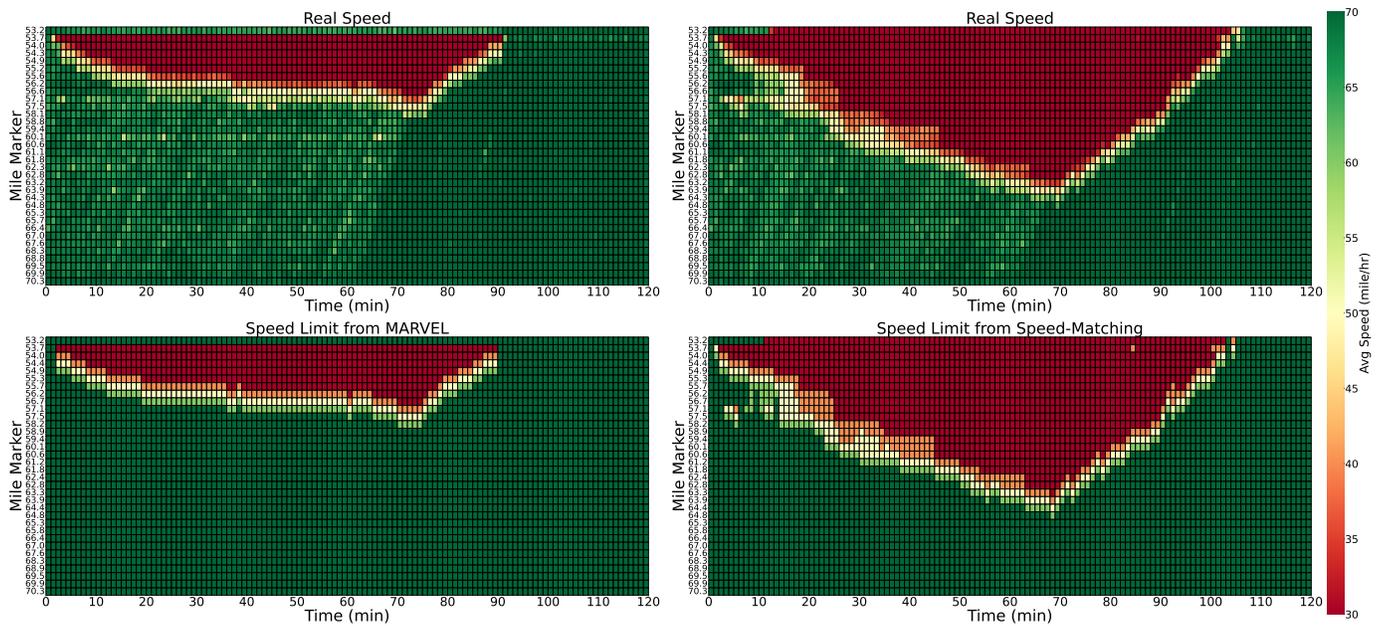}
        \label{fig:subfig_b}
    }
    \vfill
    \subfloat[Scenario C]{
        \includegraphics[width=\textwidth]{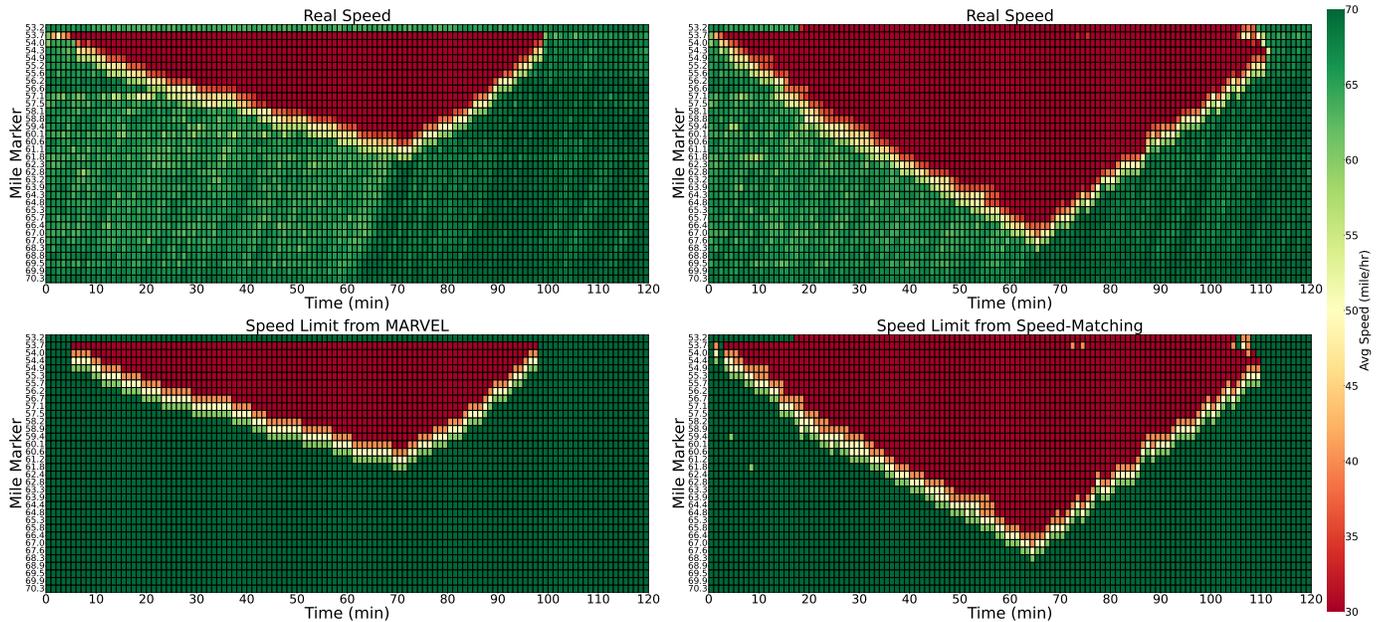}
        \label{fig:subfig_c}
    }
    \caption{The time-space diagram of Scenario B (top) and C (bottom). The left column shows the scenario under MARVEL-MAPPO control and right column shows the scenario under Speed-Matching control. (Upper row: the average traffic speed. Lower row: the control outputs from the algorithms. \textit{Note: the figure is trimmed in time to show the congestion better.})}
    \label{fig:scenario_b_c}
\end{figure*}

To understand how the learned policy from MARVEL-MAPPO behaves under varying traffic demand and compliance rates, we first present the time-space diagram of MARVEL-MAPPO along with the Speed-Matching algorithm, which is currently deployed on I-24. Figure~\ref{fig:scenario_a} and~\ref{fig:scenario_b_c} present the time-space diagram of traffic speed and the corresponding speed limits generated by each algorithm in testing scenarios A, B, and C, respectively. In scenario A, we observe two instances of recurrent congestion caused by ramp merging, a significant deviation from the training scenario which has only one such instance. MARVEL-MAPPO generalizes effectively to this new context, with agents posting low-speed limits for congested areas, high-speed limits for free-flowing areas, and intermediate speed limits for the congested tail to mitigate abrupt braking. For scenarios B and C, which exhibit higher compliance rates than the training scenario, the agents adapt well, managing longer congestion queues and making appropriate decisions. Importantly, in all testing scenarios, the agents cooperatively generate a smooth slow-down profile for the transition zone therefore improving safety by reducing speed variation. This successful inter-agent cooperation suggests the robustness of MARVEL framework and especially the correctness of our problem formulation in Section~\ref{method}, which takes into account both temporal and spatial dimensions of decision-making.

Compared with the speed-matching approach, MARVEL-MAPPO can improve traffic mobility significantly while maintaining a smooth slow-down speed profile to improve safety for the transition zone. Figure~\ref{fig:scenario_a} shows that in scenario A, the Speed-Matching approach is sensitive to the traffic oscillations and it will introduce unnecessary slowdowns which are harmful to the overall traffic mobility. On the contrary, MARVEL exhibits a more stable behavior when confronting complex traffic conditions. In scenarios B and C, which is reflected in Figure~\ref{fig:scenario_b_c}, we can observe that MARVEL-MAPPO has a significantly less congestion queue and duration than speed-matching, which emphasizes the advantage of MARVEL-MAPPO in mobility improvement even with a high compliance rate. Furthermore, the speed-matching approach results in the phenomenon of congestion propagating downstream while MARVEL-MAPPO does not introduce such unwanted behavior. The underlying reason for the suboptimal mobility performance of the speed-matching approach appears to be the simple rule-based logic, which makes it difficult to achieve an optimal solution under varying traffic conditions. In contrast, by applying MARVEL-MAPPO, the agents cooperate to maximize the safety and mobility rewards therefore leading to better performance. These results also emphasize the robustness of MARVEL-MAPPO with respect to varying compliance rates, which implies that it potentially may be applied to different deployment sites where drivers have different compliance backgrounds.


Moreover, it is worth highlighting that the number of agents has been increased to 34 in all testing scenarios, compared to 8 agents utilized during the training phase. The successful scalability for the number of agents can be ascribed to several key factors: the homogeneous training setting that enhances computational efficiency, the effective state-action-reward design, and the adoption of CTDE that ensures the agent can decide by itself during execution.

Finally, the implementation of MARVEL-IPPO and MARVEL-MAPPO-IAM also results in cooperative control among agents, although MARVEL-IPPO does not perform well from the perspective of adaptability reward term, as discussed in the following subsection. The time-space diagram of No Control and MARVEL-IPPO for testing Scenarios A, B, and C are presented in the Appendix. MARVEL-MAPPO-IAM is omitted as it has very similar results to MARVEL-MAPPO except that MARVEL-MAPPO-IAM guarantees the step-down rules.

\subsection{Testing Performance: Quantitative Analysis} 
  We conducted 10 stochastic experiments for each control method in each scenario to evaluate their performance. The variability in results across experiments is primarily influenced by the stochasticity in driver behavior parameters. Therefore, the experiments are more realistic since human drivers have varying behaviors in the real world.

\begin{figure}
    \centering
    \includegraphics[width=0.5\columnwidth]{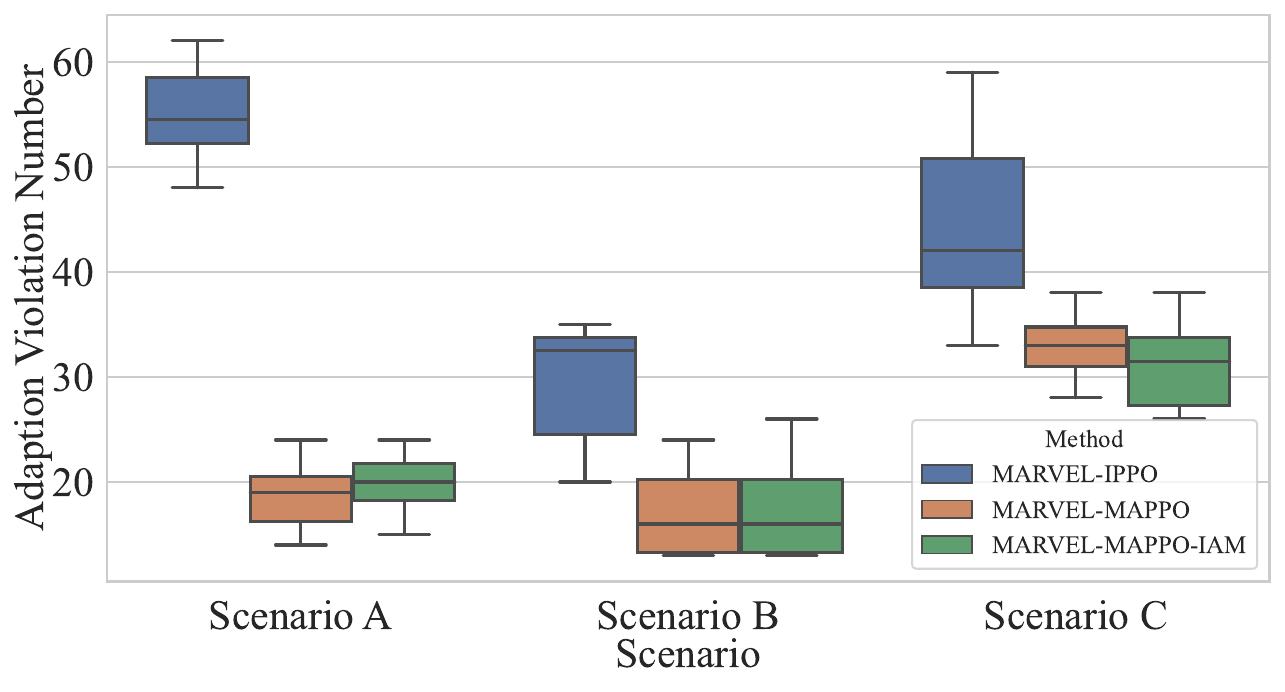}
    \caption{The boxplot of adaption violation number.}
    \label{adap_vio_counts_boxplot}
\end{figure}

\begin{figure}
    \centering
    \includegraphics[width=0.5\columnwidth]{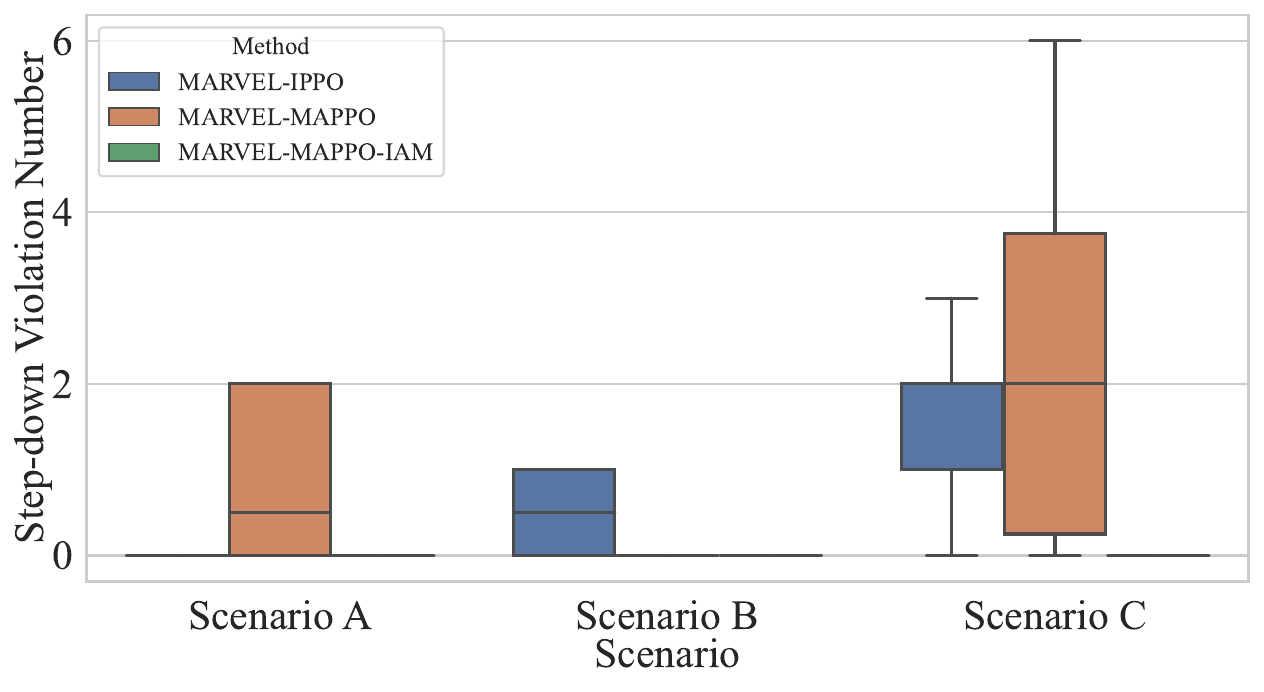}
    \caption{The boxplot of step-down rule violation number.}
    \label{sd_vio_counts_boxplot}
\end{figure}

\begin{figure*}
    \centering
    \includegraphics[width=\columnwidth]{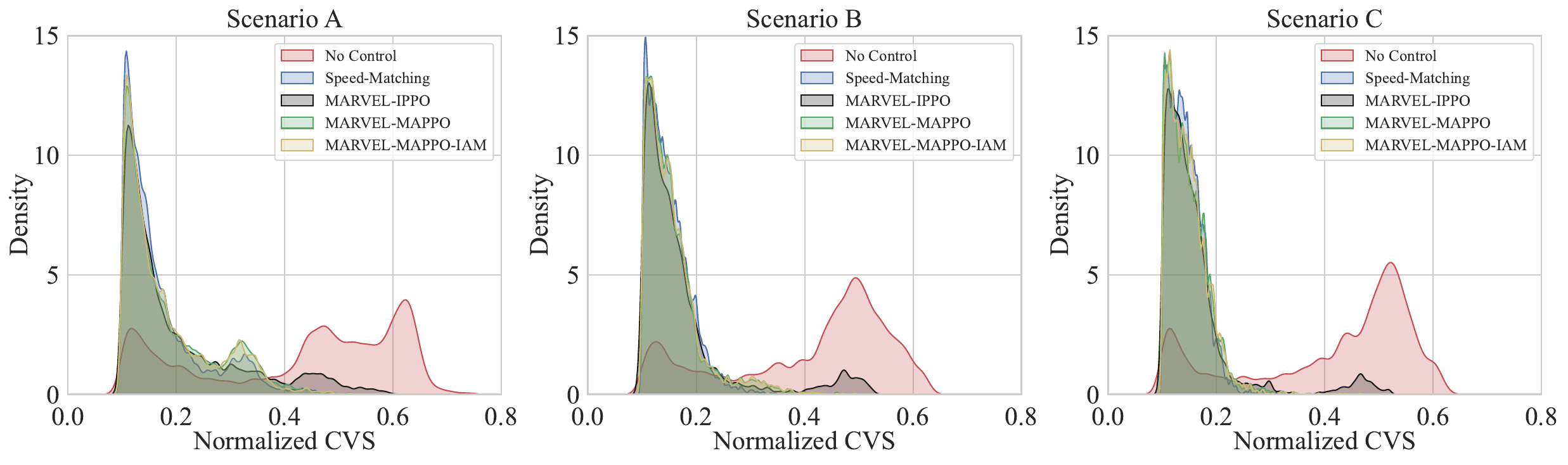}
    \caption{The distribution plot of the normalized CVS of each method for scenario A, B, and C.}
    \label{cvs_distribution}
\end{figure*}

\begin{figure*}
    \centering
    \includegraphics[width=\columnwidth]{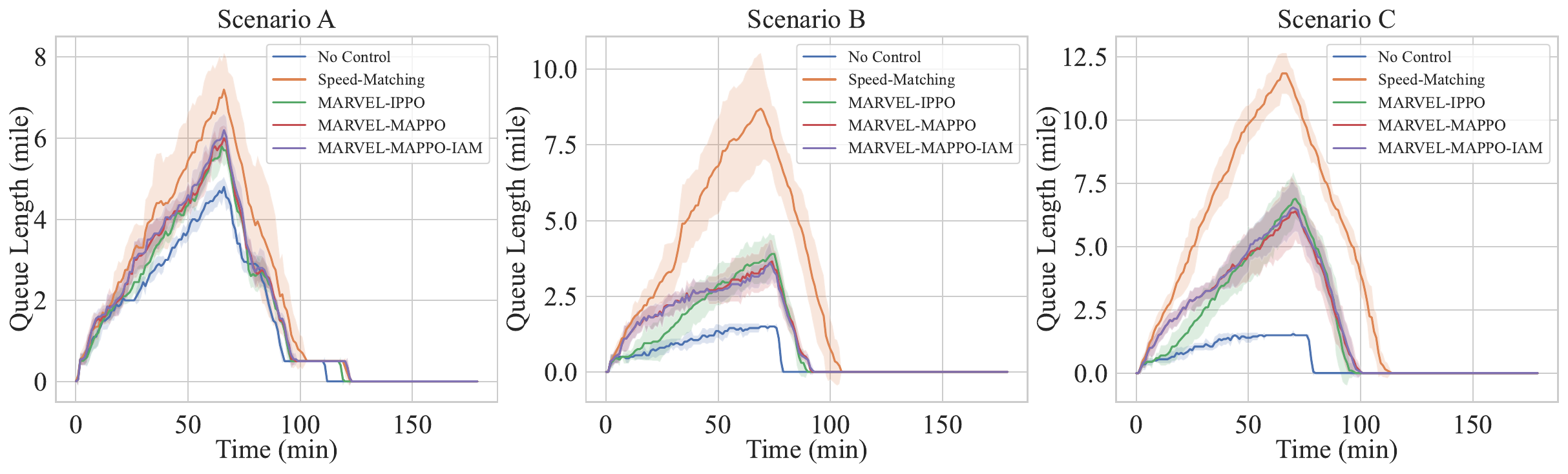}
    \caption{The time-series plot of queue length of each method for scenarios A, B, and C.}
    \label{queue_length}
\end{figure*}

We first present how MARVEL-based methods perform from the RL design perspective, i.e., how they behave in terms of adaptability and step-down rules. 
Figure~\ref{adap_vio_counts_boxplot} displays the boxplot of adaption violation for MARVEL-based methods under three scenarios. It can be observed that both MARVEL-MAPPO and MARVEL-MAPPO-IAM have much fewer adaption violations than MARVEL-IPPO across all scenarios. Along with the time-space diagram presented in the Appendix, we can infer that MARVEL-IPPO does not capture the congested state and agents post the maximum speed limit where drivers are not able to speed up. This policy will lead to two undesired results. First, as observed from the time-space diagram, the policy will fail to generate a slow-down speed profile and will not be able to improve safety. Second, this deficiency will potentially lead to distrust from road users concerning the control strategy. Figure~\ref{sd_vio_counts_boxplot} displays the boxplot of step-down violation number for MARVEL-based methods. It can be observed that both MARVEL-IPPO and MARVEL-MAPPO could violate the step-down rules sometimes. This makes sense from the RL perspective as the reward function itself cannot guarantee the avoidance of certain behaviors. However, with the help of invalid action masking, the step-down rule violation problem can be fully resolved as demonstrated by the figure. Therefore, MARVEL-MAPPO-IAM can fulfill the operational constraints in real-world deployment.

Figure~\ref{cvs_distribution} presents the distribution of normalized CVS values for the No Control baseline and all control methods across all scenarios. It can be observed that the No Control baseline has high densities for large CVS values indicating higher safety risks while the rest of the control methods can reduce CVS values in all scenarios. However, among all control methods, the Speed-Matching, MARVEL-MAPPO, and MARVEL-MAPPO-IAM have comparable safety performance while MARVEL-IPPO introduces another density peak for large CVS values. This unsatisfactory behavior can be traced back to the poor performance of MARVEL-IPPO in adaptability, which leads to unsuccessful slow-down speed profile generation, as we discussed above. Therefore, MARVEL-IPPO will sometimes lead to abrupt braking from vehicles and increase the safety risk. Additionally, Figure~\ref{cvs_distribution} verifies that MARVEL-MAPPO-IAM has a very similar safety performance with MARVEL-MAPPO indicating that there is no negative effect on traffic safety by involving the invalid action masking.

Figure~\ref{queue_length} presents the time series of queue length for the No Control baseline and the other control methods across all scenarios. It can be observed that in all scenarios, the No Control baseline has the minimum queue length, the Speed-Matching algorithm has the longest queue and the MARVEL-based methods have a performance in between. The result from the No Control baseline aligns with our expectations because given our problem setting of no capacity drop, traffic mobility cannot be improved through VSL control. Therefore, the No Control baseline will result in the shortest queue length. However, the MARVEL-based methods can significantly reduce the queue length and the congestion duration when compared with the Speed-Matching algorithm and this performance advantage is more compelling when the compliance rate is at a middle and high level. It's interesting to see that MARVEL-IPPO has a relatively shorter queue length at the early period of the congestion than other MARVEL-based methods. This can still be attributed to the fact that MARVEL-IPPO fails to capture the congestion state in the early period and does not generate the slow-down speed profile to slow down the vehicles entering into the congestion tail, while other MARVEL-based methods can do that. Note that here MARVEL-MAPPO-IAM also has a very similar mobility performance with MARVEL-MAPPO. Table~\ref{statistics} provides the mean and standard deviation for the above-mentioned metrics over all scenarios.

In summary, both the rule-based Speed-Maching algorithm and MARVEL-based methods demonstrate a significant reduction in speed variation, thereby enhancing overall safety. However, the MARVEL-based methods outperform the benchmark rule-based algorithm in mobility metrics across all testing scenarios. This performance advantage is more significant as the compliance rate increases. The enhanced mobility performance of MARVEL can be attributed to its reward design, which assigns considerable importance to the mobility term. Additionally, MARVEL induces more stable behavior and is not adversely affected by minor traffic oscillations, a common issue that impairs the performance of the Speed-Matching algorithms. It should be noted that despite the observed increase in queue length under MARVEL-based methods compared with the No Control baseline, a potential reduction in the number of primary or secondary incidents may contribute to an overall enhancement in mobility metrics. Among MARVEL-based methods, MARVEL-IPPO has a poor performance in capturing the congestion state, therefore, it induces failures in slowing down the upstream vehicles and increases the safety risks. Compared with MARVEL-MAPPO, MARVEL-MAPPO-IAM guarantees the step-down rule while achieving identical safety and mobility performance, which suggests it is an ideal choice in all benchmarks.

\begin{table*}
\centering
\caption{Mean and standard deviation of metrics with No Control, Speed-Matching and MARVEL-based control methods under all testing scenarios.}
\label{statistics}
\begin{tabularx}{ 0.7\columnwidth}{cccccc}
\toprule
\shortstack{Scenario \\ \text{ }} & \shortstack{Method\\ \text{ }} & \shortstack{Adaption \\ Vio. Num.} & \shortstack{Step-down \\ Vio. Num.}  & \shortstack{CVS\\ \text{ }} & \shortstack{Max Queue \\ Length}\\ 
\midrule 
 & NoControl & - & - & 0.43 $\pm$ 0.01 & 4.8 $\pm$ 0.2\\
  \cmidrule{2-6}
 & Speed-Matching & - & - & \textbf{0.17 $\pm$ 0.01} & 7.2 $\pm$ 0.9\\ 
A & MARVEL-IPPO  & 55.2 $\pm$ 4.1 & \textbf{0} & 0.21 $\pm$ 0.00 & \textbf{5.8 $\pm$ 0.4}\\
 & MARVEL-MAPPO & \textbf{19.1 $\pm$ 3.8} & 0.9 $\pm$ 0.9 & \textbf{0.18 $\pm$ 0.00} & \textbf{6.0 $\pm$ 0.3} \\
 & MARVEL-MAPPO (IAM) & \textbf{20.3 $\pm$ 3.6} & \textbf{0} & \textbf{0.18 $\pm$ 0.01} & 6.2 $\pm$ 0.4\\
\midrule 
 & NoControl & - & - & 0.41 $\pm$ 0.02 & 1.5 $\pm$ 0.0\\
 \cmidrule{2-6}
 & Speed-Matching & - & - & \textbf{0.15 $\pm$ 0.00} & 8.7 $\pm$ 1.8\\ 
B & MARVEL-IPPO  & 31.3 $\pm$ 8.6 & 0.5 $\pm$ 0.5 & 0.18 $\pm$ 0.01 & 3.9 $\pm$ 0.6\\
 & MARVEL-MAPPO & \textbf{17.1 $\pm$ 4.0} & \textbf{0} & \textbf{0.15 $\pm$ 0.01} & \textbf{3.7 $\pm$ 0.7}\\
 & MARVEL-MAPPO (IAM) & \textbf{17.4 $\pm$ 4.5} & \textbf{0} & \textbf{0.15 $\pm$ 0.00} & \textbf{3.6 $\pm$ 0.7}\\
\midrule 
 & NoControl & - & - & 0.41 $\pm$ 0.02 & 1.6$ \pm$ 0.2 \\
 \cmidrule{2-6}
 & Speed-Matching & - & - & \textbf{0.14 $\pm$ 0.00} & 11.9 $\pm$ 0.8\\ 
C & MARVEL-IPPO & 46.4 $\pm$ 12.6 & 2.0 $\pm$ 1.0 & 0.17 $\pm$ 0.01 & 6.9 $\pm$ 0.8\\
 & MARVEL-MAPPO & \textbf{33.0 $\pm$ 3.0} & 2.0 $\pm$ 1.9 & \textbf{0.15 $\pm$ 0.00} & \textbf{6.4 $\pm$ 1.1}\\
 & MARVEL-MAPPO (IAM)  & \textbf{31.1 $\pm$ 4.1} & \textbf{0} & \textbf{0.15 $\pm$ 0.00} & \textbf{6.5 $\pm$ 1.0}\\
\bottomrule
\end{tabularx}
\end{table*}

\begin{figure*}[h]
    \centering
    \includegraphics[width=\columnwidth]{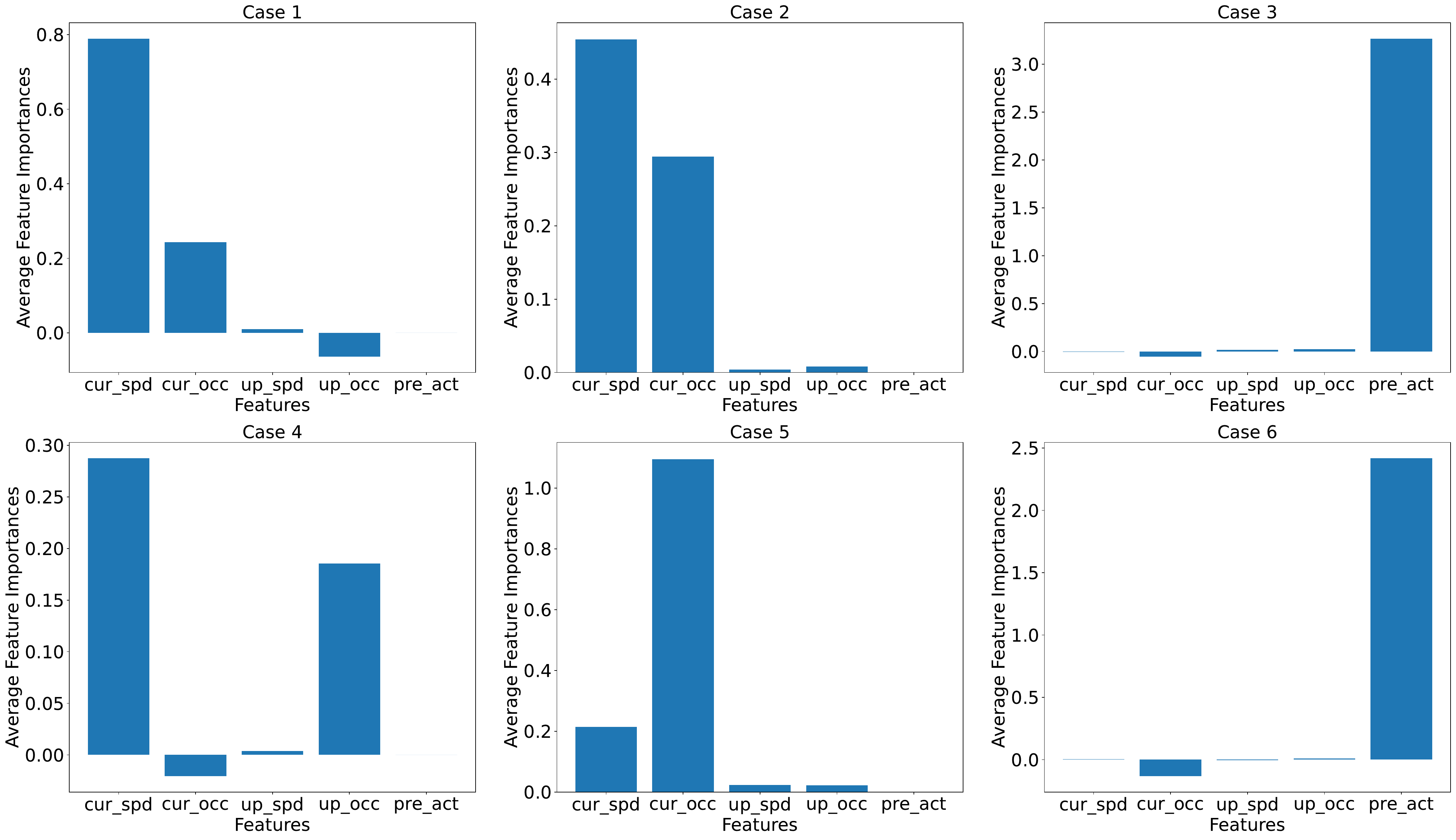}
    \caption{The average feature importance of six different traffic scenarios.}
    \label{explanability}
\end{figure*}

\subsection{Explainability Analysis}
To further understand the agents' decision-making and evaluate the effectiveness of the state space design for each agent, we conduct an explainability analysis based on the integrated gradients method~\cite{sundararajan2017axiomatic}. Integrated gradients is a machine learning interpretability method that calculates the importance of each feature by integrating the model's gradients from a baseline to the input. To understand the change in agents' decision-making under different traffic scenarios, we define the following six cases:
\begin{itemize}
    \item \textbf{\textit{Case 1:}} Agent selects 70 at time $t-1$ and 30 at time $t$.
    \item \textbf{\textit{Case 2:}} Agent selects 40 at time $t-1$ and 30 at time $t$.
    \item \textbf{\textit{Case 3:}} Agent selects 50 at time $t-1$ and 40 at time $t$.
    \item \textbf{\textit{Case 4:}} Agent selects 30 at time $t-1$ and 70 at time $t$.
    \item \textbf{\textit{Case 5:}} Agent selects 30 at time $t-1$ and 40 at time $t$.
    \item \textbf{\textit{Case 6:}} Agent selects 40 at time $t-1$ and 50 at time $t$. 
\end{itemize}

For each of the aforementioned cases, we collect five samples from the interaction dataset of MARVEL-MAPPO. We employ the state information at $t-1$ as the baseline, and the state information at $t$ as the input. The chosen action at time $t$ serves as the target variable. We then compute the mean attribution for each input feature, as visualized in Figure~\ref{explanability}. Our analysis reveals that significant features exhibit a positive correlation with the target, especially in the direction from the baseline to the input states. More specifically, agents tend to rely more heavily on current speed and occupancy when transitioning from a free-flow state or a transitional state to a congested state, and vice versa. We note that the agent depends on both the local traffic information and the upstream traffic information to decide to recover the speed limit as shown in case 4. Additionally, when situated in transition zones, agents place greater emphasis on the actions of preceding agents for decision-making. This analytical investigation offers valuable insights into the decision-making processes of agents across diverse traffic conditions, as facilitated by the MARVEL framework. Furthermore, the study serves as empirical evidence for the efficacy of our state design.


\subsection{Response to Empirical Data}
\begin{figure*}[h]
    \centering
    \includegraphics[width=\textwidth]{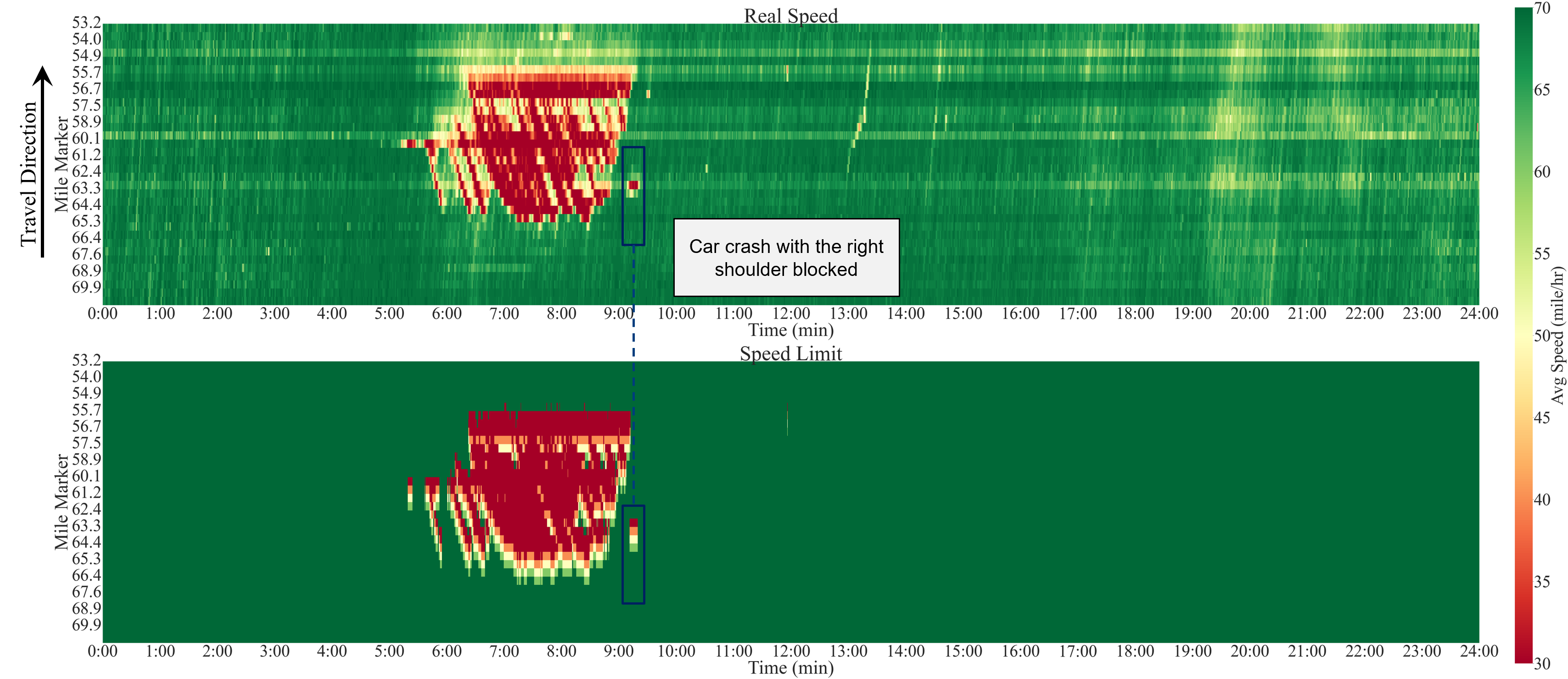}
    \caption{The time-space diagram of I-24 westbound on 01-08-2024. (Upper: traffic speed recorded by RDS units. Lower: the control behavior from MARVEL-MAPPO-IAM. Note that the agents respond to the non-recurrent congestion caused by a right-shoulder blockage event.)}
    \label{deployment}
\end{figure*}

To further illustrate the generalizability and deployment potential of the policy learned through the MARVEL framework, we assess the behavior of MARVEL-MAPPO-IAM in an open-loop manner using empirical data from I-24 Smart Corridor as described in Section~\ref{experiment}. This data, which the agents have never encountered during training, provides a robust test bed to evaluate the learned policy's responsiveness to actual traffic scenarios without affecting the real traffic.

Figure~\ref{smartcorridor} reveals that multiple RDS units may be situated between two consecutive VSL controllers.  To utilize this data, each VSL agent is assigned proximate downstream RDS units, up to the jurisdiction of the next downstream agent. Among the assigned RDS units, we identify the most critical one to guide each agent's decision-making. The selection criterion is based on occupancy and speed metrics: if the occupancy across all units is consistently high or low, the unit with the minimum speed is prioritized; otherwise, the unit with the maximum occupancy is chosen.

By applying MARVEL-MAPPO-IAM in real-time to the empirical RDS data, we observe and record the agents' decision-making process. Figure~\ref{deployment} shows the traffic speed and the agents' decision-making from I-24 westbound on January 8, 2024. The traffic speed time-space diagram indicates that the morning peak-hour traffic occurs around 5:30 AM and then multiple stop-and-go traffic waves can be observed, which is a typical traffic pattern on I-24 that has also been detected by I-24 MOTION~\cite{gloudemans202324}, a state-of-the-art traffic monitoring system. Although there is no specific stop-and-go wave generated in the simulation, MARVEL-MAPPO-IAM can respond efficiently to several early traffic waves and provide a smooth slow-down speed profile for the upstream traffic to improve safety. During the off-peak hour time, the agents are posting the maximum speed limit to improve traffic efficiency. It is worth highlighting that there is a non-recurrent congestion caused by a car crash around 9:10 AM and MARVEL-MAPPO-IAM responds reasonably well in this scenario even though it was never trained for it.

While this evaluation does not involve a closed-loop control scenario, where the agents' decisions directly influence traffic conditions, it represents an essential step in assessing whether the actions generated by the learned policy are acceptable when it is extracted from simulation and directly applied to real-world empirical data. Our finding demonstrates that MARVEL-MAPPO-IAM can identify the real-world congestion state, generate a smooth slow-down speed profile to warn the upstream drivers and define a faster speed when the conditions allow. This encouraging result showcases the readiness of more advanced testing phases for MARVEL-based methods, including measuring the sim-to-real gap from the perspective of MDP and potential pilot implementations.


\section{Conclusion}
\label{conclusion}

Most of the deployed VSL control methods so far are rule-based because of their explainable and simple nature. However, rule-based methods are often reactive and could induce unnecessary travel time delays. In this article, we introduce MARVEL, a MARL framework for large-scale VSL control for considerations of real-world deployment on I-24 in Nashville, TN, USA. Specifically, we consider three requirements for learning-based VSL to be eventually deployable, namely, scalability, generalizability, and feasibility. In addition, we consider three objectives for large-scale VSL control including adaptability to congestion traffic state, safety, and mobility. 

We compare MARVEL-based methods with the Speed-Matching algorithm currently deployed on I-24, in a microscopic traffic simulator. We apply the MAPPO algorithm and a derivative of MAPPO with invalid action masking (MAPPO-IAM), both under the MARVEL framework.
Our results show that MARVEL-MAPPO and MARVEL-MAPPO-IAM  generalize across a range of traffic conditions (varying demands and driver compliance rates), and scale up to typical roadway corridors. The evaluation of the two algorithms spans diverse traffic scenarios and demonstrates a significant reduction of congestion queue length while preserving the safety achievement compared to the Speed-Matching algorithm. The results suggest that MARVEL-MAPPO-IAM achieves an identical traffic safety and mobility performance with MARVEL-MAPPO, and at the same time, it guarantees the step-down rules, which makes it promising for real-world deployment. To validate the real-world deployment capability of MARVEL, we test the response of MARVEL-MAPPO-IAM policy to one-day real data collected from I-24 roadside sensors. The time-space diagram of the response suggests that the agents can correctly capture the congestion state and post the minimum speed limit (adaptability), coordinate with each other to generate a smooth slow-down speed profile for upstream traffic (safety), and post the maximum speed limit when the traffic condition allows (mobility). 

There are three future works that we are interested in. Firstly, we will focus on testing MARVEL-MAPPO-IAM on a real deployment on I-24 and measuring the mobility and safety performance on the freeway. Secondly, we will quantify the drivers' response to VSL control on I-24, which provides us insights into the sim-to-real gap. Thirdly, we will collect the interaction dataset for potential offline training after our deployment. 


\section*{Acknowledgments}
The contents of this report reflect the views of the authors, who are responsible for the facts and accuracy of the information presented herein. This work is supported by a grant from the U.S. Department of Transportation Grant Number 693JJ22140000Z44ATNREG3202. However, the U.S. Government assumes no liability for the contents or use thereof. The authors are grateful to  Caliper for technical support on the TransModeler micro-simulation software used in this work.

\bibliographystyle{IEEEtran}
\bibliography{ref}

\appendix

\begin{figure}[htbp]
    \centering
    \includegraphics[width=0.5\columnwidth]{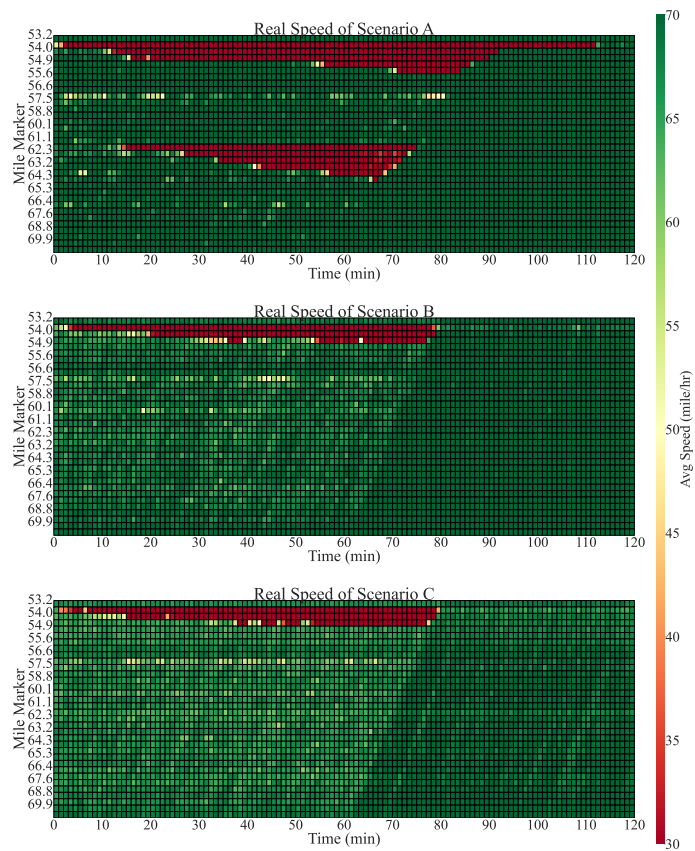}
    \caption{The average traffic speed time-space diagram of Scenario A (top), B (middle) and C (bottom) when under No Control baseline. \textit{Note: the figure is trimmed in time to better show the congestion.})}
    \label{no_control_ts}
\end{figure}

\begin{figure}[htbp]
    \centering
    \includegraphics[width=0.5\columnwidth]{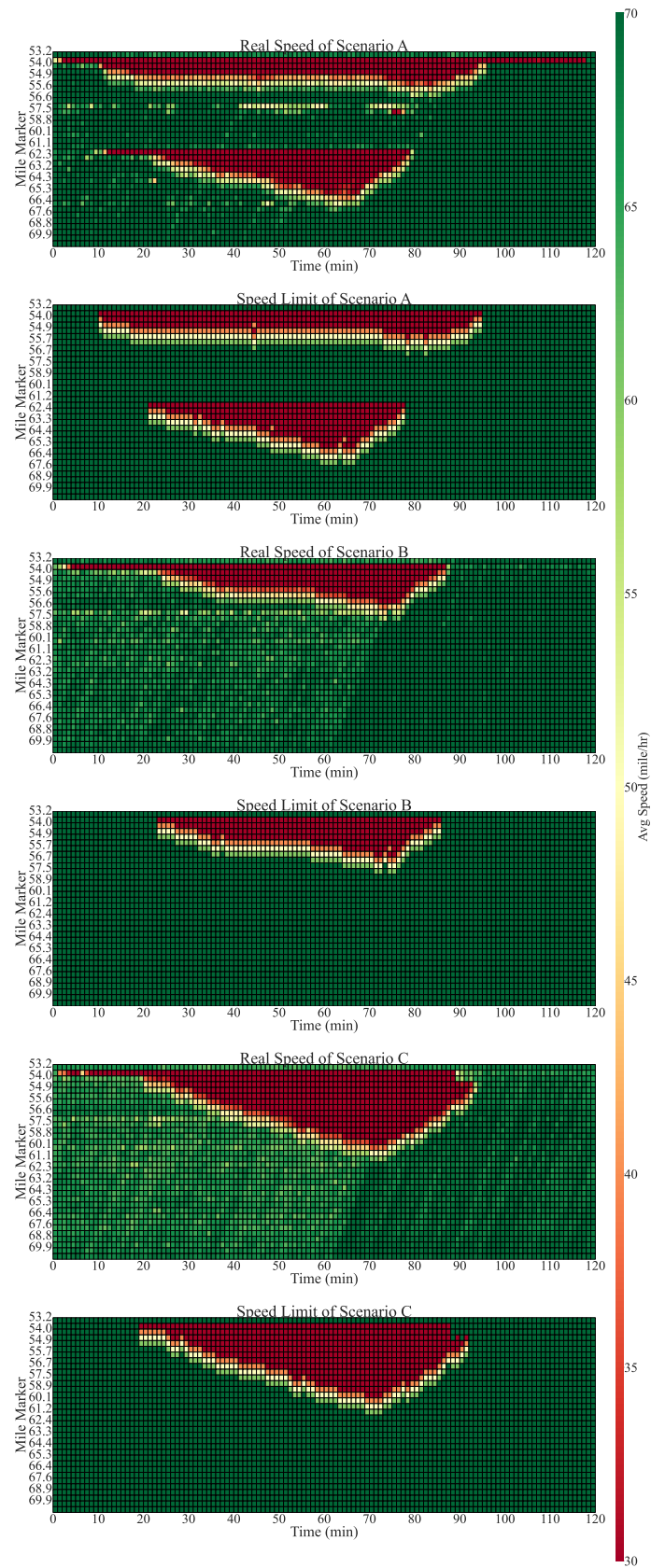}
    \caption{The time-space diagram of Scenario A(top), B (middle) and C (bottom) when under MARVEL-IPPO control. (In each scenario, the top row represents the average traffic speed and the bottom row represents the control outputs from the algorithm. \textit{Note: the figure is trimmed in time to better show the congestion.})}
    \label{ippo_ts}
\end{figure}

\vfill

\end{document}